\documentclass{aa}

\usepackage{graphicx}
\usepackage{xcolor}
\usepackage{subcaption}
\usepackage[colorlinks,linkcolor=blue,citecolor=blue,urlcolor=blue]{hyperref}
\usepackage{cleveref}

\usepackage{txfonts}

\begin{document}

   \title{Differences in the physical properties of satellite galaxies within relaxed and disturbed galaxy groups and clusters}

   \author{F. Aldás
          \inst{1,2}
          \and
          Facundo A. Gómez\inst{2,3}
          \and
          C. Vega-Martínez\inst{4}
          \and
          A. Zenteno\inst{5}
          \and
          Eleazar R. Carrasco\inst{6}
          }

   \institute{INAF – Osservatorio Astronomico di Trieste, Via G. B. Tiepolo 11, 34143 Trieste, Italy. \\
              \email{franklin.aldas@inaf.it}
         \and
         Departamento de Astronomía, Universidad de La Serena, Avenida Juan Cisternas 1200, La Serena, Chile.\\
         \and
             Instituto de Investigación Multidisciplinar en Ciencia y Tecnología, Universidad de La Serena, Raúl Bitrán 1305, La Serena, Chile.\\
        \and
        Facultad de Ingeniería y Arquitectura, Universidad Central de Chile, Av. Francisco de Aguirre 0405, La Serena, Chile.\\
        \and
        Cerro Tololo Inter-American Observatory/NSF NOIRLab, Casilla 603, La Serena, Chile.\\
        \and
        International Gemini Observatory/NSF NOIRLab, Casilla 603, La Serena, Chile. \\
             }

   \date{Received August 05, 2024; accepted May 27, 2025}

  \abstract

   {Galaxy groups and clusters are the most massive collapsed structures in the Universe. They assemble hierarchically through the successive mergers of smaller systems. These dense environments play a crucial role in driving the evolution and morphological transformation of their galaxies.}
   { The dynamical state of groups and clusters can affect the properties of their galaxy populations. Our aim is to characterise the distribution of galaxies' colour, specific star formation rate, quenched galaxy fraction, and gas availability in galaxies bounded to groups and clusters and to examine how these properties relate to the dynamical state of their host environments. }
   {We used the most massive halos (M > 10$^{13}$ M$_\odot$) in the Illustris TNG100 simulation and separated the sample into two categories: relaxed and disturbed halos. This classification was done based on the offset between the position of the brightest cluster galaxy (BCG) and the centre-of-mass of the gas. Subsequently, we classified their galaxy populations into red and blue galaxies using a threshold derived from a double Gaussian fit to their colour distribution. Additionally, we distinguished between star-forming and quenched galaxies by applying a threshold defined as one dex below the interpolated star formation main sequence. }
   {Our findings reveal differences in physical properties such as colour, star formation rates, and gas availability among satellite galaxies bound to interacting clusters compared to relaxed clusters. Disturbed clusters exhibit more blue, star-forming galaxies than their relaxed counterparts. This discrepancy in the fraction of blue and star-forming galaxies can be attributed to higher gas availability, including hot, diffuse, and condensed gas in satellite galaxies in disturbed clusters compared to relaxed ones. Furthermore, our study shows that during cluster mergers, there are two crucial phases; at the beginning of interaction, there is an important boost in the star formation rate, followed by a suppression as the cluster reaches the equilibrium state. }
   {}


   \maketitle
%
\section{Introduction}
Galaxy groups and clusters are the most massive gravitationally bound structures in the Universe, containing up to thousands of luminous galaxies~\citep{Abell1957, Abell1989}. According to the $\Lambda$-CDM ($\Lambda$-cold dark matter) model, the formation of structures in the Universe followed a hierarchical process.  Initially, small overdensities in the primordial Universe collapsed and later developed into larger systems, through merging and accretion with other smaller structures~\citep{Cole2000, White1978, White1991}.
Environmental mechanisms such as ram-pressure stripping (RPS), tidal interactions, and pre-processing drive galaxy evolution within groups and clusters. RPS can remove loosely bound gas from infalling galaxies, sometimes triggering bursts of star formation or active galactic nucleus (AGN) activity before quenching sets in \citep{Balogh2000, Cole2000, Somerville2008, Ayromlou2019, Dressler1980, Vulcani2018, Roberts2022}. Tidal forces strip stars, gas, and dark matter, contributing to intracluster light~\citep{Moore1999, Gnedin2003, Mastropietro2005}.  Moreover, galaxies often undergo preprocessing in group environments prior to cluster infall, with quenching efficiency increasing in more massive hosts ~\citep{Zabludoff1996, Mcgee2009, Bahe2019, Pallero2019}. 

Due to the influence of the described environmental effects, galaxies in clusters exhibit markedly different physical properties compared to those in the field. One of the most fundamental observations linking galaxy properties and environment is the morphology-density relation, whereby early-type galaxies are more prevalent in high-density environments~\citep{Dressler1980, Butcher1984, Kauffmann2004, Vulcani2011, Sazonova2020}. In these environments, galaxies typically show suppressed star formation activity~\citep{Lewis2002, Kauffmann2004, Fassbender2014, Peng2010, Woo2013, Finn2023}, reduced AGN activity ~\citep{Galametz2009, Bufanda2017, Poggianti2017, Marshall2018}, and lower gas content~\citep{Stroe2015, Poggianti2017, Cairns2019} compared to their field counterparts. Furthermore,~\citet{Skibba2009} found that more massive groups tend to host redder satellite populations. Additionally, the fraction of star-forming satellites decreases with increasing halo mass, a trend that becomes more pronounced at lower redshifts~\citep{Wang2018}. Finally, while the overall shape of the galaxy luminosity function remains broadly similar across environments, its composite form varies due to differences in galaxy type distributions. Cluster environments tend to exhibit a steeper faint-end slope and a more extended bright-end tail, driven primarily by an excess of dwarf spheroidal and bright ellipticals galaxies~\citep{Dressler1978, Garilli1999, Goto2002, Christlein2003, Valotto1997, Binggeli1988}.

Properties of galaxies in clusters also show a clear dependence on redshift. At higher redshifts, clusters host a lower fraction of early-type galaxies \citep{Postman2005} and a higher fraction of actively star-forming galaxies than their low-redshift counterparts. This trend indicates a gradual decline in star formation activity within cluster galaxies over cosmic time \citep{Margoniner2001, Poggianti2006, Saintonge2008, DeLucia2024}. This evolution is accompanied by an increase in blue galaxies at higher redshift \citep{Butcher1978, Goto2003, Haines2015, Poggianti2006, DeLucia2004, Saintonge2008, Aldas2023, Wen2018}, and a corresponding decline in AGN activity as the Universe evolves \citep{Galametz2009, Bufanda2017}. Redshift evolution is also evident in the cluster galaxy luminosity function \citep{Haines2009, Sarron2018, To2020, Lin2006}. The characteristic luminosity ($L^*$) increases with redshift, reflecting a combination of declining star formation and ongoing stellar mass growth over time \citep{Crawford2009, To2020, Norberg2001}. While many galaxy properties exhibit well-established redshift trends, the evolution of the metal abundance remains less certain. Some studies suggest that the average metallicity remains roughly constant at around 0.3 times the solar value \citep{Donahue1998, Tozzi2003}, whereas others report a decrease of up to 50\% between $z = 0.1$ and $z = 1$ \citep{Maughan2008, Anderson2009}.

Despite the assumption of equilibrium often used in cosmological analyses, a substantial fraction, between 30 and 70$\%$, of galaxy clusters are currently in an unrelaxed dynamical state~\citep{Hou2012}.  An unrelaxed cluster -- also referred to
as a disturbed or dynamically active system -- is a galaxy cluster that has not yet reached dynamical equilibrium. In such clusters, the main components (dark matter, hot intracluster gas, and galaxies) are still responding to recent or ongoing processes such as mergers, mass accretion, or interactions with the cosmic web~\citep{Hou2012, Brunetti2008, Nurgaliev2013, Gouin2021}. 
A merging system of galaxy clusters typically undergoes three principal stages over a timescale of approximately one gigayear \citep{Wilber2019}. These include a pre-merging phase, during which the clusters begin their infall and the intracluster medium (ICM) start to interact; a merging phase, characterised by the core passage; and a post-merging phase, in which the ICM gradually relaxes as shocks and turbulence dissipate \citep{Wilber2019}. The dynamical state of a cluster can be inferred from multi-wavelength observations. Radio halos and relics are typically associated with disturbed clusters~\citep{Brunetti2008}, while cool cores are indicative of relaxation~\citep{Bauer2005}. X-ray morphological parameters such as centroid shifts, concentration, and power ratios have been widely used as relaxation indicators~\citep{Nurgaliev2013, Campitiello2022}. Cosmological simulations provide alternative metrics, including the virial ratio, centre-of-mass offsets, and substructure mass fractions~\citep{Bett2007, DeLuca2021, Zhang2022}.

Several studies have reported systematic differences in galaxy properties between relaxed and disturbed clusters. Galaxies in disturbed clusters tend to have a bluer and more diverse galaxy population than the relaxed ones ~\citep{Aldas2023, Wang1997, Cortese2004, Hou2012, Cava2017}. In this sense, \citet{Kelkar2023} also reported that post-merger clusters, despite sharing a red galaxy population with relaxed clusters, retain a significant number of blue galaxies. However, they did not find any direct evidence of merger-driven galaxy processing. Regarding star formation activity, there is no consensus on whether disturbed clusters enhance or suppress star formation. For example, ~\citet{Cohen2014}  and \citet{Yoon2020} reported that galaxies in interacting clusters exhibit higher star formation rates (SFRs) than those in non-interacting systems, while ~\citet{Yoon2019} found that merger clusters trigger star formation specifically in the bars of cluster galaxies. In contrast, other studies, such as those by \citep{Kleiner2014} and \citep{Shim2011}, found no significant enhancement in star formation, and some even suggest that interacting clusters may suppress it~\citep{Deshev2017}. AGN activity is closely linked to the dynamical state of galaxy clusters, with evidence indicating that cluster mergers can enhance AGN triggering. For example, \citet{Bilton2020} demonstrated that merging clusters host younger AGN populations compared to those in more dynamically relaxed systems. Moreover, ~\citet{Yoon2019} showed that interacting clusters have a fraction of 1.5 times higher fraction of barred disc-dominated galaxies in comparison with no interacting ones. Finally, relaxed galaxy clusters exhibit a pronounced central peak in iron abundance ($\text{r} < 0.2 \; \text{R}_{500}$), whereas non-relaxed clusters show a more uniform Fe distribution due to core disruption and metal redistribution during mergers~\citep{Urdampilleta2019}.

Beyond direct environmental effects, large-scale correlations such as galactic conformity have emerged as important probes of galaxy evolution. This phenomenon, manifested as a correlation between the star formation of central galaxies and that of their satellite or neighbouring galaxies, has been robustly identified across cosmic time and environments. Using deep near-infrared data at intermediate to high-redshift (0.4<z<1.9), \citet{Hartley2015} found that passive satellites preferentially reside around passive centrals, with no significant redshift evolution, suggesting that conformity is not solely driven by halo mass but may also arise from environmental or feedback processes that impact both centrals and their satellites. Similarly, \citet{Kawinwanichakij2016} reported a significant conformity signal in the redshift range 0.6<z<1.6 with satellites around quiescent centrals exhibiting higher quiescent fractions compared to those around star-forming centrals. This trend remains  significant after accounting for halo mass differences, suggesting a connection between star formation activity of central galaxies and the fraction of quenched satellites. On the theoretical side, \citet{Ayromlou2022} analysed conformity in galaxy formation models (L-GALAXIES, IllustrisTNG, and EAGLE) and observational data from SDSS and DESI, finding conformity signals extending to several megaparsecs. Their results suggest that physical processes such as RPS, feedback, and the influence of backsplash and fly-by galaxies near massive halos contribute to the conformity signal.
Furthermore, the dynamical state of clusters has been linked to their cosmic web connectivity, a measure of the number of filaments connected to the cluster. While some studies report a positive correlation between dynamical activity and connectivity~\citep{Gouin2021}, others find no significant trend~\citep{Santoni2024}. Still, this connectivity likely plays a role in cluster growth and environmental pre-processing.

In this paper, we unveil the influence of the galaxy cluster's dynamical state on the physical properties of their galaxy population, such as colour, star formation, metallicity, and gas availability. We also investigate the physical mechanisms responsible for causing these changes. This work was done using a fully cosmological simulation Illustris-TNG that includes a model to describe galaxy formation and evolution.

\section{Cosmological simulations}
\label{Sec:Simulaciones}
\begin{figure}
\centering
\includegraphics[width=\columnwidth, trim={3cm, 2.8cm, 2.0cm, 3cm}, clip]{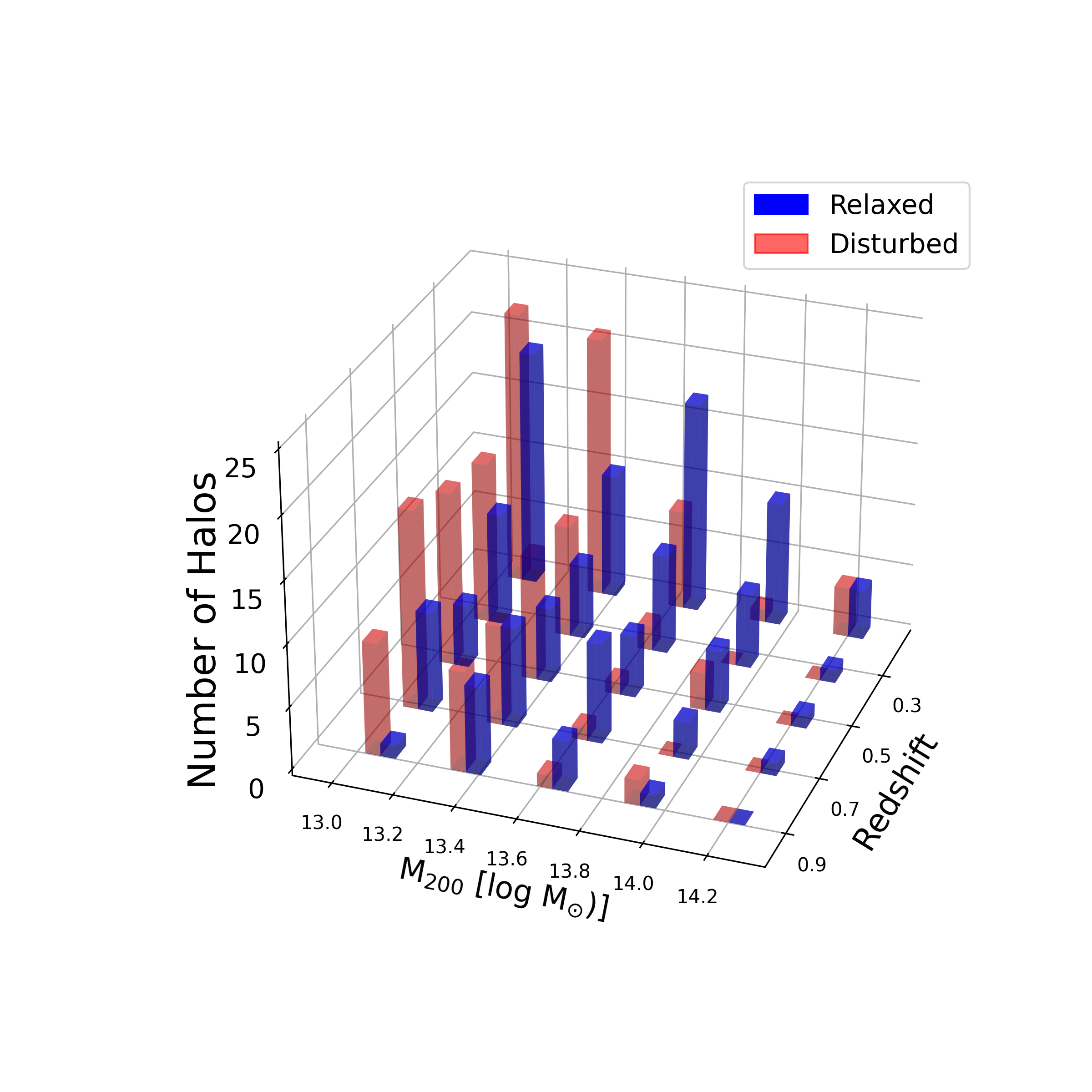}
\caption{Halos selected for this study in function of their virial mass ($\text{M}_{200}$) in the x axis and their redshift (z) in the y axis. In blue we have the selected relaxed clusters and in red the disturbed sample. }
\label{Haloes_distribution}
\end{figure}

In this project, we use the Illustris TNG cosmological magnetohydrodynamical simulations, which are a series of large simulations aimed at exploring the formation and evolution of galaxies within the $\Lambda$-Cold Dark Matter model framework~\citep{Pillepich2018, Springel2018, Naiman2018, Nelson2018, Marinacci2018}. Illustris TNG was run with the AREPO code, a flexible code for hydrodynamical cosmological simulations that solves Poisson's equation in a fully dynamical unstructured mesh with a finite-volume method  ~\citep{Weinberger2020}. These simulations include inhered models from Illustris simulations incorporating non-linear physical processes driving galaxy formation and evolution, such as star formation, stellar evolution, chemical enrichment, and gas recycling~\citep{Vogelsberger2014, Genel2014, Sijacki2015}. Illustris TNG also has updated models for the growth and feedback of supermassive black holes, galactic winds, and metal abundance evolution~\citep{Weinberger2017, Pillepich2017model}. 
This model exhibits a good agreement with observational data regarding the abundance of molecular and atomic gas in the IllustrisTNG, as compared to the abundances derived with the 21-cm Hydrogen line using the Giant Metre-wave Radio Telescope ~\citep{Rhee2018, Diemer2019}. 

This suite of simulations has three different box sizes: $50$ Mpc,  $100 $ Mpc, and $300$ Mpc per side, hereafter TNG50, TNG100, and TNG300, respectively. Each box has different mass resolutions for Dark Matter (DM) and baryonic particles. The groups and clusters in the Illustris-TNG simulations were identified using the Friend-of-Friends algorithm with a linking length of 1 h$^{-1}$kpc, 0.5 h$^{-1}$kpc, and 0.2h$^{-1}$kpc for TNG300, TNG100, and TNG50, respectively~\citep{Springel2018}. Their corresponding subhalos were identified using the SUBFIND algorithm ~\citep{Springel2001, Dolang2009}. The TNG300 simulation has a volume 27 times larger and contains approximately 20 times more halos than TNG100 ~\citep{Pillepich2018}. However, TNG100 offers a mass resolution that is one order of magnitude better than TNG300. Specifically, TNG100 contains  $1820^3$  DM and an equivalent number of baryonic particles with a mass resolution of $7.5\times 10^6$ M$_\odot$ and $1.4\times 10^6$ M$_\odot$, respectively. For this work, we have chosen to use the TNG100 to balance the number of detected structures and a good resolution. TNG100 has more than 2000 halos with masses greater than $10^{12}$ M$_{\odot}$, and the most massive object detected at redshift $z=0$ has a mass of $3.8\times 10^{14}$ M$_\odot$. 

The Illustris TNG simulations reproduce notably well several observed statistical properties of the galaxy population. These properties include the clustering of galaxies as a function of stellar mass, colour, star formation, and redshift~\citep{Springel2018}, the galaxy colour bimodality~\citep{Nelson2019}, and the stellar mass functions from $z=4$ to today~\citep{Pillepich2018}.  
Illustris TNG  uses the cosmological parameters from Planck observations~\citep{Planck2015}, as follows: $\Omega_b=0.0486$, $\Omega_{m}=0.3089 $,  $\Omega_\Lambda=0.6911$, and $H_0=67.74$ km s$^{-1}$ Mpc$^{-1}$. 

All halo and subhalo properties used in this study are derived using the Friends-of-Friends (FoF) algorithm for halos ~\citep{Davis1985} and the SUBFIND algorithm for the subhalos~\citep{Springel2001, Dolang2009}. These algorithms group particles based on their spatial proximity and binding energy. 

\section{Sample selection}
\label{Sec:Sample}

 \begin{figure}
		\centering
		\includegraphics[width=\columnwidth, trim={1cm 0.5cm 2.5cm 2cm}, clip]{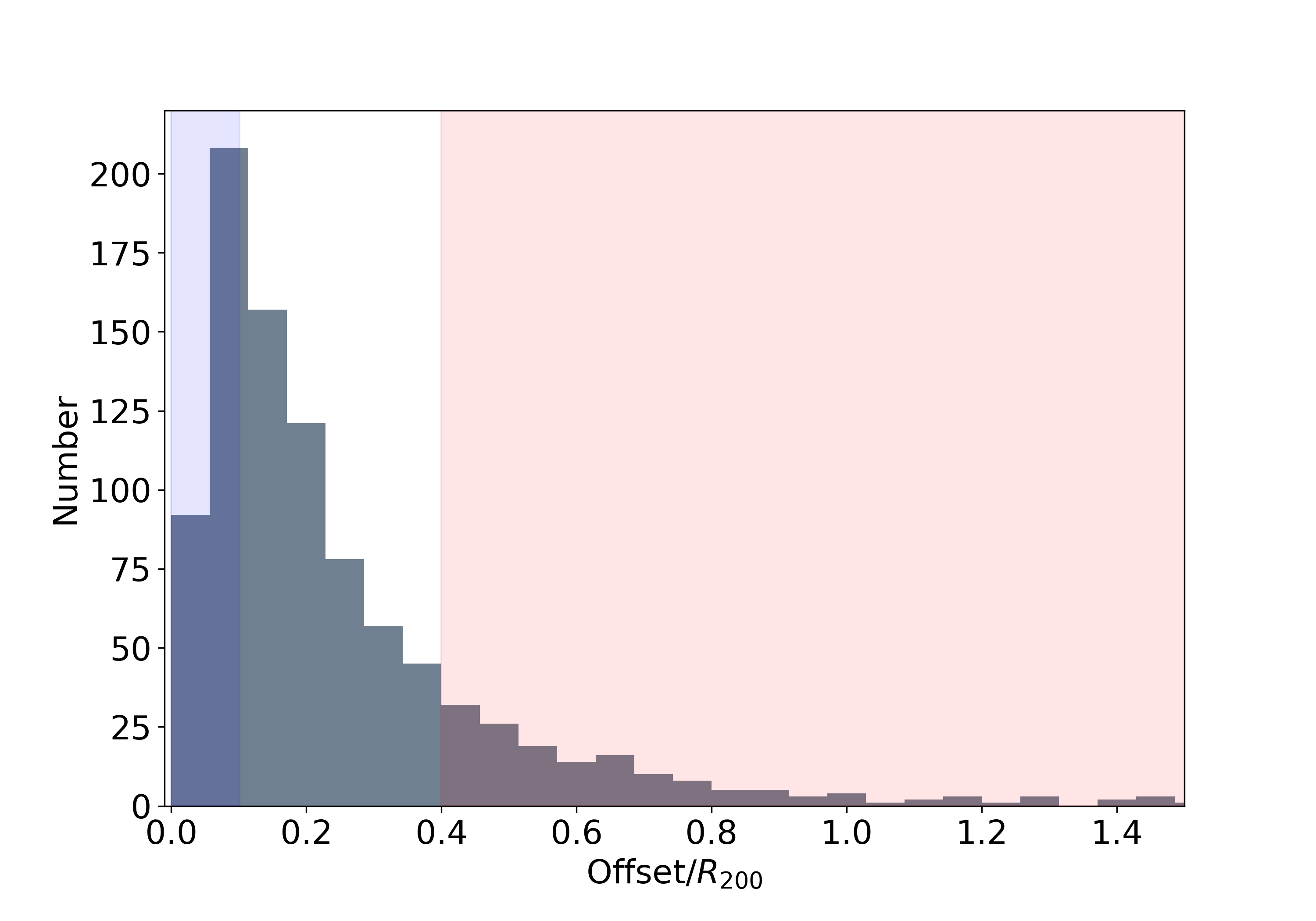}
		\caption{Offset between the position of the BCG and the centre-of-mass of the gas used as an indicator of the dynamical state of groups and clusters. In the blue shadow area, the selected 153 most relaxed groups and clusters between $0.1 \leq z \leq 1$,  and in the red shadow area, the selected 153 disturbed ones. }
		\label{Fig:Offset_hist}
\end{figure}
 In this paper, we analyse the properties of galaxy populations in groups and clusters. Our focus will be on structures with masses greater than M$_{200}\geq 10^{13}$ M$_{\odot}$ containing at least ten galaxies per halo. Where the viral radius, R$_{200}$, is defined as the radius within which the mean enclosed mass density is 200 times the critical density of the Universe, and the total mass within this radius is the virial mass denoted M$_{200}$. This paper considers as galaxies subhalos with stellar mass greater than  $10^9$ M$_\odot$, resolving each galaxy with more than 800 stellar particles. We look for clusters and groups of galaxies independently in different simulation snapshots, assuming that the dynamical state of the structures changes over time. We select halos from seven different redshifts, between $0.1$ and $1$.  The redshifts used are 0.1, 0.2, 0.3, 0.4, 0.5, 0.7, and 1. Following ~\cite{Zenteno2020} and \cite{Aldas2023}, who studied the differences in properties of cluster galaxies at low and high-redshift, we subdivide our sample of galaxies into two redshift bins. The first is a low-redshift bin from $0.1 \leq z< 0.5$ and a high-redshift bin with halos between $0.5\leq z\leq1$. 
 \begin{figure*}
		\centering
		\includegraphics[width=\linewidth, trim={5cm, 2cm, 4cm, 4cm}, clip]{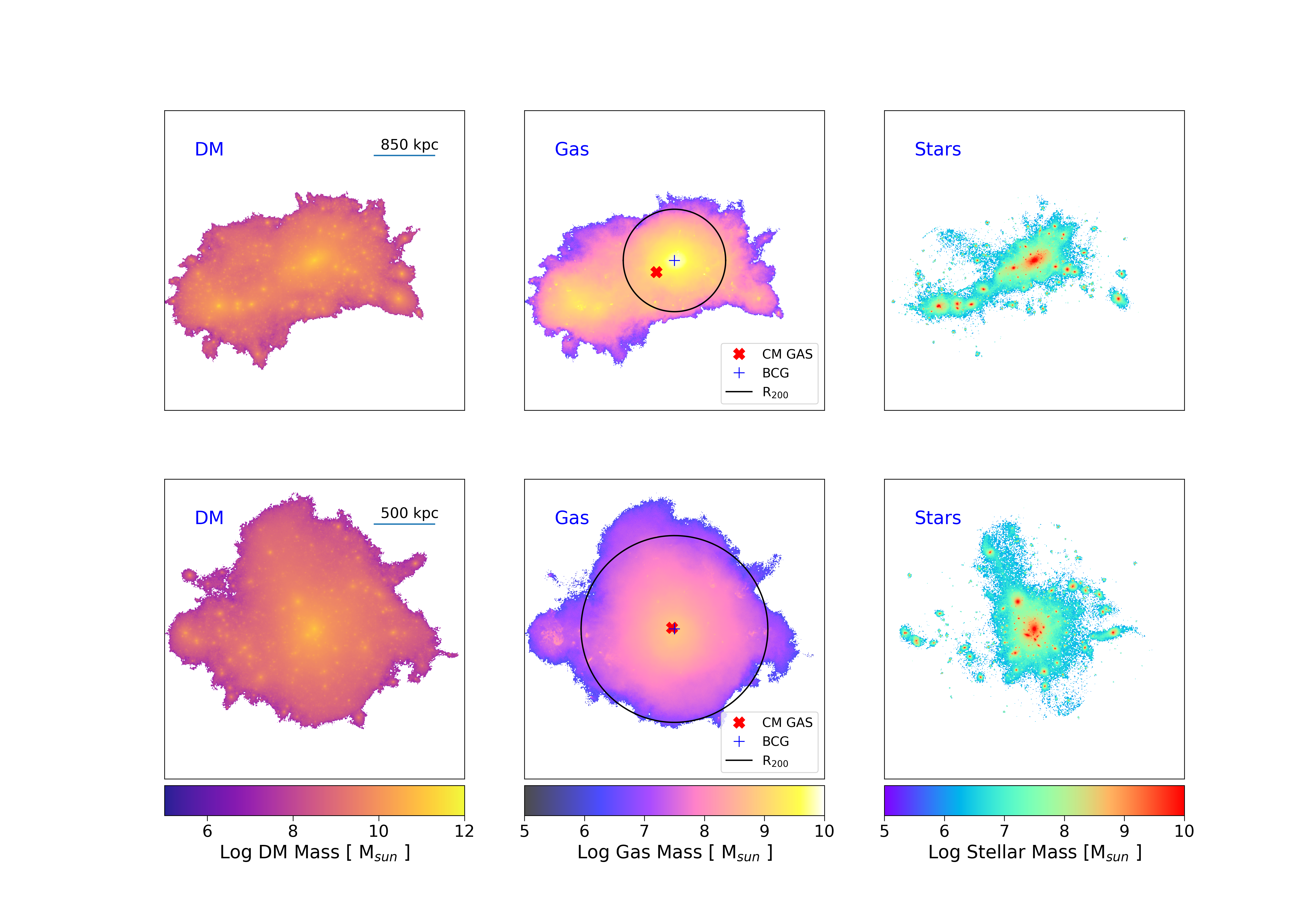}
		\caption{In the top panels, the dark matter (left), gas (centre) and stars particles (right) mass distribution for the most massive disturbed halo at redshift $z=0.1$ of the Illustris TNG simulations. We also show the BCG's position and the centre-of-mass of gas cells in the central panel; we can see that the position of the centre-of-mass is located around $0.4 \text{R}_{200}$ measured from the position of the BCG. In the bottom panels, similar to the top panels, the distribution of the different components for the most massive relaxed halo at the same redshift.    }
		\label{Fig:Examples}
\end{figure*}
In Figure~\ref{Haloes_distribution}, we present the selected groups and clusters' mass and redshift distribution. The redshift is presented on the x axis, and the virial mass is on the y axis.  The mean temporal separation between the used snapshots is 1.3 Gyrs, which represents around half the free-falling time for two collapsing structures with masses of $10^{14}\; \text{M}_{\odot}$ initially separated 1 Mpc. This ensures that the dynamical state of clusters will be different from one selected snapshot to another.   
 
\label{DynamicalState}

To study the dynamical state of galaxy clusters, following ~\cite{Zenteno2020}, we separated our sample of groups and clusters into relaxed and disturbed categories using an observationally motivated parameter: the 3D offset between the position of the BCG and the position of the centre-of-mass of the gas. Following conventions in cosmological simulations, the BCG is defined as the galaxy located at the potential minimum of a group or cluster, provided it does not lie within the R$_{200}$ of a more massive halo~\citep{Pillepich2018}. This potential minimum corresponds to the position of the most bound particle within the halo and also defines the centre of the group or cluster.  On the other hand, the centre-of-mass of the gas is computed using all gas cells bound to the halo. We consider that a cluster is disturbed if the offset between the BCG and the centre-of-mass of the gas (D$_{\text{BCG-CM}}$) is larger than $0.4\times \text{R}_{200}$.  We use the position of the centre-of-mass of the gas as a proxy for the SZ effect centroid. Our proposed methodology aligns closely with the approach outlined by~\cite{DeLuca2021}. In their study, they incorporate various indicators for the dynamical state, including the 3D offset between the centre-of-mass of the cluster considering all particle types and the position of the highest density peak. When considering the centre-of-mass computed across all particle types in our groups and clusters and imposing the same threshold of $0.4 \times \text{R}_{200}$, we find that  $85\%$ of the groups and clusters meet the disturbed classification criterion in both methodologies.

 In the simulation, we found 640 groups and clusters with virial masses greater than $10^{13}$ M$_{\odot}$ within  the selected seven snapshots. From those clusters, we identified 153 as disturbed structures. To make a fair comparison of the cluster galaxies' physical properties between relaxed and disturbed clusters, we also selected the 153 most relaxed clusters. These are the ones with the lowest D$_{\text{BCG-CM}}$ ($<0.01$ R$_{200}$). Our criterion for identifying relaxed clusters is more stringent compared to the approach suggested by ~\citet{Cui2017} for the centre-of-mass offset, leading us to focus solely on the most relaxed clusters. In Figure~\ref{Fig:Offset_hist}, we present a histogram of the distribution of $D_{BCG-CM}$ in units of R$_{200}$. The blue and red shadowed areas show the selected relaxed and disturbed clusters, respectively. These selection criteria allow us to maximise the differences between the two samples. As a result, clusters that are in an intermediate relaxation state, which are located in between the shaded areas of Figure~\ref{Fig:Offset_hist}, have not been considered in this study.

As was previously described, all selected halos have an identified BCG, and we assume that relaxed clusters contain only a single central galaxy. In contrast, for disturbed clusters, we define a secondary BCG as the most massive galaxy located at a distance greater than the offset between the cluster’s centre-of-mass and the position of the primary BCG. If the unrelaxed halo corresponds to an ongoing merger, the secondary BCG typically represents the central galaxy of the infalling substructure. Otherwise, it is interpreted as the most dominant galaxy associated with the perturbation responsible for the cluster's dynamical disturbance. In all unrelaxed halos selected in our sample, the secondary BCG corresponds to the second most massive galaxy in the system. Finally, satellite galaxies are defined as all non-BCG subhalos bound to the host,  with a stellar mass of at least $\text{M}_* > 10^9 \text{M}_\odot$. It is important to note that not all subhalos host stellar components; some consist exclusively of dark matter particles and are excluded on our analysis.

To assess the relative mass of the secondary BCG compared to other galaxies in the cluster, we examined the ratio of the secondary BCG's mass to that of the most massive non-BCG galaxy in the group. Our analysis shows that in 93\% of cases, the secondary BCG is at least twice as massive as the most massive non-BCG galaxy in the cluster. Furthermore, in 62\% of the unrelaxed halos, the secondary BCG is at least five times more massive than the most massive non-BCG galaxy. These findings indicate that, in most cases, a clearly dominant galaxy can be identified as the secondary BCG.

As is shown in Table \ref{table:number_clusters}, the relaxed clusters have 4442 galaxies, while the disturbed ones have slightly larger substructures, corresponding to 4494 galaxies. 
Each sub-sample, i.e. disturbed and relaxed, is further subdivided into a low ($0.1\leq z<0.5$) and high ($0.5\leq z \leq 1$) redshift sub-sample. As a result, we end up with four distinct galaxy groups.
The final number of galaxy clusters and the number of their member galaxies in each subgroup is given in Table~\ref{table:number_clusters}. Each subgroup has at least 63 clusters and groups with at least 1633 bounded galaxies. 

Comparing the mass differences between the selected relaxed and disturbed galaxy groups and clusters, we find that, despite applying a mass threshold in our selection, relaxed halos are, on average, 7\% more massive than disturbed clusters. Moreover, when comparing low- and high-redshift bins, clusters in the low-redshift bin are, on average, 41\% more massive than those in the high-redshift bin. This significant mass–redshift dependence is consistent with the hierarchical model of structure formation, in which clusters grow over time through the continuous accretion of matter and mergers. The observed increase in mass from high- to low-redshift reflects the cumulative nature of these growth processes as the Universe evolves.

A similar analysis of the  BCGs reveals that in the low-redshift bin those are consistently 40\% more massive than their counterparts in the high-redshift bin. Furthermore, BCGs in relaxed clusters tend to be approximately 30\% more massive than those in disturbed systems. This suggests that, like their host clusters, BCGs continue to grow in mass over time—likely through mechanisms such as galaxy mergers and the accretion of smaller satellites. The parallel trends between overall cluster mass and BCG mass highlight the co-evolution of clusters and their central galaxies.

\begin{table}
\centering
\caption{Number of clusters in the selected sample in each analysed subgroup. We divided the sample into low and high-redshift bins and relaxed and disturbed clusters. In parenthesis, we also show the number of bounded galaxies. }
\label{table:number_clusters}
\begin{tabular}{ccc}
\hline
z & \# Rel. clust. (Gal.) & \# Dist. clust. (Gal.) \\
\hline
$0.1<z<0.5$ & 90 (2809) & 80 (2682) \\
$0.5\leq z<1$ &  63 (1633) & 73 (1802)\\
Total & 153 (4442) & 153 (4494)\\
\hline
\end{tabular}
\end{table}

The left, middle, and right columns of Figure~\ref{Fig:Examples} show the DM, gas, and stellar mass distribution of an example of disturbed (top panels) and relaxed (bottom panel) cluster, respectively. 
In the central panels, the blue, '+'  mark represents the position of the most bound particle, which traces the location of BCG. The red 'x' mark represents the position of the CM of the gas distribution,  and the black circle represents the virial radius (R$_{200}$) of the halo centred in the BCG position. The disturbed cluster (top panels) has a halo with  $\text{M}_{200}=2.7\times 10^{14} \; \text{M}_{\odot}$ at redshift $z=0.1$. We can see that the position of the CM of the gas has an offset of $0.42 $ \text{R}$_{200}$ from the position of the BCG due to another massive structure that is infalling in the central halo. On the other hand, in the bottom panels, we present a halo with a mass of $\text{M}_{200}= 3.28\times 10^{14} \; \text{M}_{\odot}$, also at redshift $z=0.1$, as an example of one relaxed cluster. We can see that, despite some small structures falling into the central halo, the position of the centre of gas remains the same as the position of the BCG, and the cluster's shape is much more symmetrical compared to the example of the disturbed one. In each example, the distribution of gas and DM mass follows similar shapes, in accordance with results by \citet{Rasia2004}, who showed that the gas and DM follow similar density profiles with different concentration values for $r> 0.1 \text{R}_{200}$.  
Finally, we can observe that the stellar mass is mainly distributed in the inner parts of the halo, where the gas density is large enough to trigger the star formation.   

\section{Galaxy populations}
\label{Section:Populations}
In this section, we analyse the colour of galaxies, establishing a threshold to divide between red and blue galaxy populations. Additionally, we classify galaxies based on their star formation activity into star-forming and quenched galaxies. We consider all galaxies bound to groups and clusters, regardless of their dynamical state. 
The impact of the dynamical state on the physical properties of satellite galaxies will be studied in Section~\ref{Sec:Differences}.

\subsection{Colour distributions}
\label{Sec:Blue/Red}
 Galaxies are classified into red and blue based on their colours, providing an insight into their dominant stellar populations. Red galaxies are characterised by a prevalence of late-type stars, while blue galaxies are dominated by early-type stellar populations. The Balmer break offers a way to identify sample sets of star-forming galaxies because its presence is common in A-type stars. The presence of A-type stars and their limited lifetime indicate recent star formation~\citep{Mihalas1967, Poggianti1997}. The Balmer break can be found by spectrography and photometry using specifically chosen colour indices. For this reason, in this paper, we define the colour of a galaxy as the difference in the magnitude between the simulated $g-$ and $r-$ SDSS broad-band filters. TNG100 simulations provide the rest frame absolute magnitudes for each galaxy computed as the summed-up luminosity of all-star particles gravitationally bounded to the subhalo in different passbands~\citep{Pillepich2018}.
 
\begin{figure}
    \centering
    \includegraphics[width=\columnwidth, trim={0.0cm, 2.5cm, 3cm, 3cm}, clip]{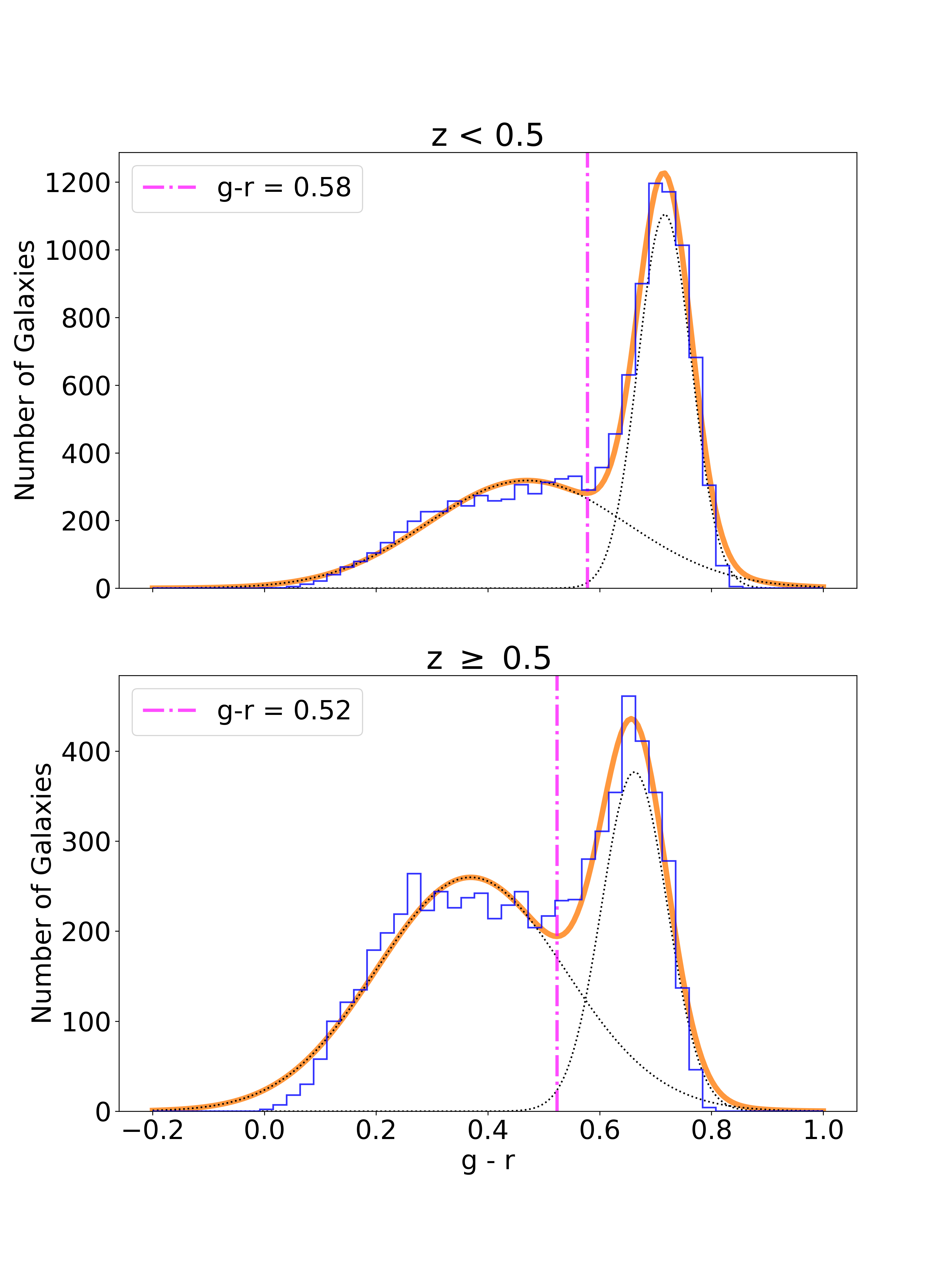}
    \caption{Galaxy colour distribution in $g - r$ (SDSS passbands in rest frame). In the top panel, we present the histogram for colours corresponding to galaxies at low-redshift $(z<0.5)$ and in the lower panel, the histogram of colour distribution for galaxies belonging to clusters at high-redshift ($0.5<z<1$). In both panels, the orange line corresponds to a double Gaussian fit, and the dotted pink line corresponds to the threshold to separate between red and blue galaxies, defined as the minimum of the colour distribution.  These thresholds are located at $0.58$, and $0.52$  for the low- and high-redshift sub-samples, respectively.  }
    \label{Fig:Double-Gaussian_color}
\end{figure}
The division into red and blue galaxies is typically done by using a threshold in the colour of the galaxies. As the threshold is expected to vary with redshift due to the Butcher-Oemler effect \citep{Butcher1984}, we define a value for low-redshift and another for high-redshift galaxies. To determine these values, we followed the pipeline proposed by \citet{Guglielmo2019}. This process consists of fitting a double-Gaussian model to the galaxy's colour distribution and using the local minimum of its distribution as the threshold. In the top panel of Figure~\ref{Fig:Double-Gaussian_color}, we show the double-Gaussian fit for galaxies in low-redshift clusters  ( $z < 0.5$), while the bottom panel shows the results for the high-redshift clusters  ( $z \geq 0.5$).  
 Figure~\ref{Fig:Double-Gaussian_color}  shows that the colour distribution for galaxies is bimodal, with two well-defined peaks corresponding to the blue and red galaxy populations~\citep{Nelson2018}. 
 The peaks in the distribution of the blue and red galaxy populations are centred at $0.46$ and $0.72$ in (g-r) for the low-redshift bin, and are located at $0.37$ and $0.66$ for the high-redshift bin. The thresholds, determined by the minimum between the two peaks, are g-r = $0.58$ for low-redshift galaxies and g-r = $0.52$ for high-redshift galaxies. We define a galaxy as red if its colour exceeds the threshold, while a galaxy is classified as blue if its colour falls below the threshold.  These values obtained for the thresholds align with the findings of \cite{Guglielmo2019}, which suggest that this value gets lower as the redshift increases.

 Figure~\ref{Fig:Double-Gaussian_color} illustrates that in the low-redshift bin, the majority of the galaxy population is made up of red galaxies, while blue galaxies are more significant in the high-redshift bin. This difference can be quantified by comparing the heights of the Gaussian curves. For low-redshift galaxies, the peak of the red population is approximately 3.5 times higher than that of the blue population. However, in the high-redshift bin, the peak of the red population is only 1.5 times higher than that of the blue population. This procedure allowed us to naturally define two populations, effectively distinguishing between blue and red galaxies.

\subsection{Star-forming and quenched galaxies}
\label{Section:Quenched_definition}

Galaxies can be classified based on their star formation activity into two categories: star-forming and quenched galaxies. Star-forming galaxies are characterised by ongoing star formation activity, while quenched galaxies have consumed, depleted or heated their gas reservoir, resulting in the cessation of star formation. 
Then, defining a single threshold to classify star forming and quenched galaxies is challenging since star-formation activity depends on the galaxy's mass and redshift. This paper follows the definition of quenched galaxies proposed by \cite{Donnari2019} and \citet{Pillepich2019}, which states that a galaxy is considered quenched if its SFR is one dex below the extrapolated star-forming main sequence (MS). \citet{Donnari2021} fitted the location of the MS in the stellar mass  versus SFR diagram for the galaxies in  the Illustris-TNG300 simulations using the relation 
\begin{equation}
    \log\Big( \frac{\langle \text{SFR}\rangle _{\text{sf-ing galaxies}}}{\text{M}_\odot yr^{-1}}\Big) = \alpha (z) \log \Big(\frac{\text{M}_{\text{stars}}}{\text{M}_{\odot}}\Big) +\beta(z),
    \label{MS_fitting}
\end{equation}
where $\alpha(z)$, and $\beta(z)$ are coefficients obtained from a iterative fitting process. The values for the coefficients are presented in Table 3 of \cite{Donnari2021}. Despite the definition adopted in this paper, our main results and conclusions remain unchanged when repeating the analysis using the widely accepted criterion from \citet{Wetzel2012}, in which galaxies are classified as quenched if their specific star formation rate (sSFR) satisfies $\log(\text{sSFR})<-11$. Moreover, as shown by ~\citet{Donnari2021} and ~\citet{Pallero2019}, the two are equivalent at low redshifts. 

In this article, SFR of each galaxy has been is calculated by summing the contributions from all gravitationally bound gas cells, corresponding to the instantaneous values available in the IllustrisTNG catalogues~\citep{Donnari2019}. Observationally, this quantity is typically inferred from flux measurements in two distinct passbands, which provide a time-averaged estimate. Although both approaches trace ongoing star formation, the instantaneous measurement captures the current activity, while a galaxy’s colour reflects its recent star formation history.

\section{Differences between relaxed and disturbed clusters}
\label{Sec:Differences}
\begin{figure}
		\centering
		\includegraphics[width=\linewidth, trim={1cm, 2.3cm, 3cm, 2cm}, clip]{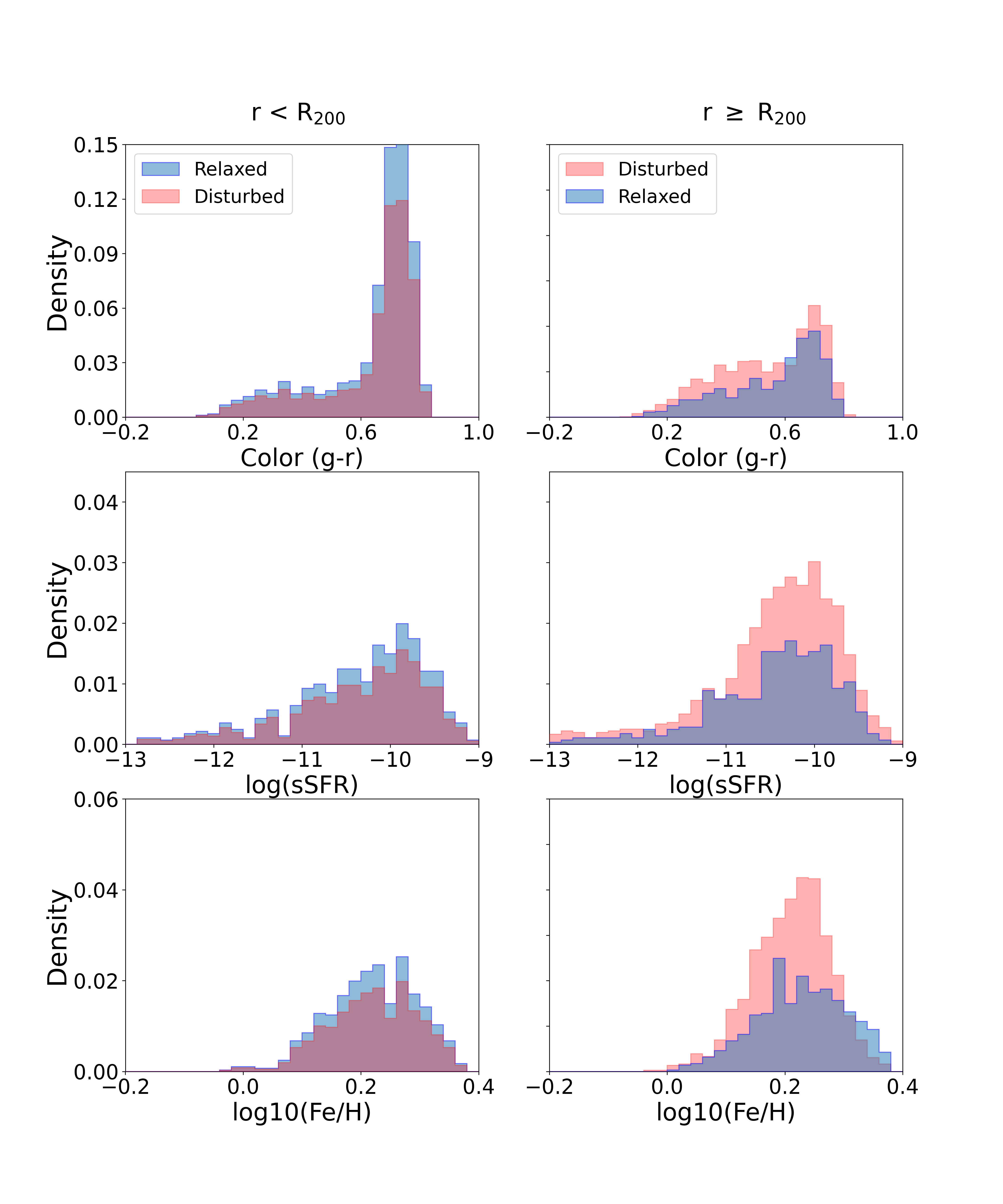}
		\caption{Differences in the physical properties (colour, sSFR and metallicity) of the satellite galaxies belonging to relaxed (blue) and disturbed (red) clusters in clusters at low-redshift bin $(0.1\leq z <0.5)$. We also separate the satellite galaxies inside $\text{R}_{200}$ (left panels) and outside $\text{R}_{200}$ (right panels). }
		\label{Fig:LR_ColorvsSM}
\end{figure}

\begin{figure}
		\centering
		\includegraphics[width=\linewidth, trim={1cm, 2.3cm, 3cm, 1.5cm}, clip]{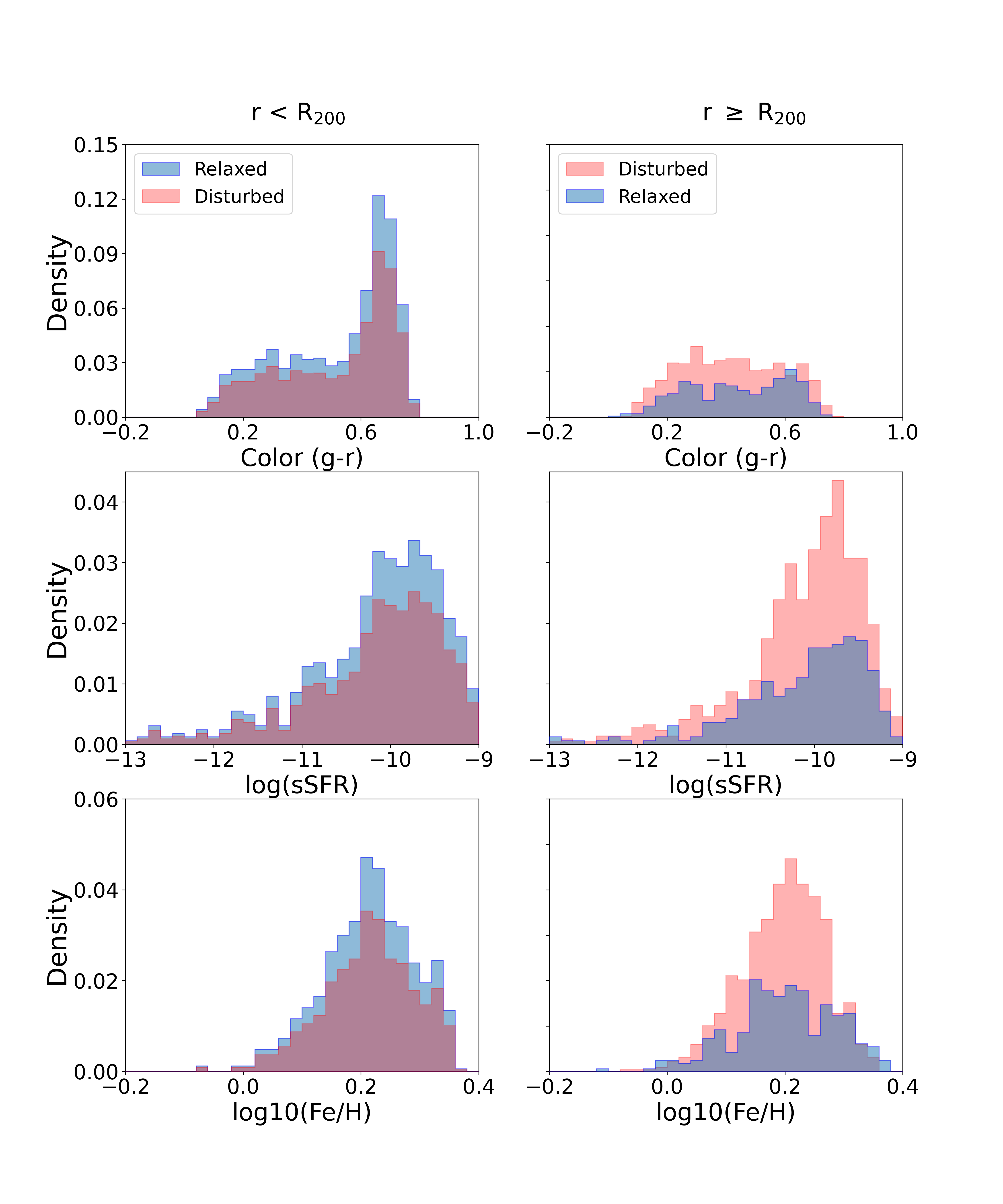}
		\caption{Differences in the physical properties (colour, sSFR and metallicity) of the satellite galaxies belonging to relaxed (blue) and disturbed (red) clusters in the high-redshift bin $(0.5\leq z \leq 1)$. We also separate the satellite galaxies inside $\text{R}_{200}$ (left panels) and outside $\text{R}_{200}$ (right panels). }
		\label{Fig:HR_ColorvsSM}
\end{figure}

In this section, we use the relaxed and disturbed clusters selected in a redshift range of $0.1\leq z \leq 1$ to study the dependence of the physical properties of galaxies on the dynamical state of their host clusters. Since properties such as galaxy colour, specific star formation rate and metallicity are dependent on redshift, i.e. they change as galaxies evolves in its environment, our sample is henceforth divided into relaxed and disturbed galaxy clusters, as well as into high- and low-redshift bins. In this context, Figure~\ref{Fig:LR_ColorvsSM} shows histograms of the galaxies' colour, sSFR, and stellar metallicity (Fe/H) for galaxies in clusters at low-redshift bin ($0.1\leq z<0.5$), while Figure~\ref{Fig:HR_ColorvsSM}, presents the same properties for galaxies bound to clusters in the high-redshift bin ($0.5\leq z \leq 1$). In those figures, the left panels present the galaxies within $\text{R}_{200}$, while the right panels present those outside $\text{R}_{200}$. In each panel, the histogram for galaxies inhabiting disturbed clusters is shown in red, while for galaxies inhabiting relaxed clusters in blue. 
All histograms are normalised by the corresponding total number of galaxies, i.e. inside and outside $\text{R}_{200}$. 

First, in both redshift bins,  we notice a larger fraction of galaxies outside the virial radius (up to 3 $\text{R}_{200}$) in the disturbed clusters compared to the relaxed ones. This is expected since disturbed clusters are interacting with other structures that also have their satellite galaxies, often located outside the main structure's $\text{R}_{200}$. In the top panels, we observe a well-defined  red sequence (RS),  centred at $ \sim 0.7$ in colour (g-r), dominating the galaxy distribution, especially in the inner regions of the clusters. However, within $\text{R}_{200}$, the fraction of galaxies in the RS is smaller in the disturbed sample with respect to their relaxed counterpart. The RS peak is also present for galaxies in the outer regions of the cluster (outside $\text{R}_{200}$)  in the low-redshift bin, but is less prominent in the high-redshift bin. The differences in the populations between red and blue galaxies are quantified using the threshold derived in Section~\ref{Sec:Blue/Red}. Our results show that, in the inner regions, 77\% (56\%) of galaxies in relaxed clusters are classified as red in the low- (high- ) redshift bin, compared to 66\% (46\%) in disturbed clusters. In the outer regions, the red fraction declines to 58\% (35\%) for relaxed clusters and 50\% (28\%) for disturbed ones.
These results show that the outer parts of clusters have a population of galaxies bluer than those within $\text{R}_{200}$, which are denser environments, and RPS is stronger, thus quenching the galaxies more efficiently. In both redshift bins, disturbed clusters exhibit a higher fraction of blue galaxies in their outer regions, indicating more recent star formation activity compared to their relaxed counterparts. These results agree with those presented by \citet{Ferrari2003}, and \citet{Lokas2020}, showing that infalling galaxies to the cluster display bluer colours. Meanwhile, the galaxies inside the $\text{R}_{200}$ become redder as they spend time interacting with the Intra Cluster Medium~\citep{Pallero2022}.  As expected, we find that early-type galaxies tend to dominate in the centre of clusters, and late-type galaxy population is more important in the outskirts of clusters~\citep{Dressler1980}. Globally, in both redshift bins, satellite galaxies in disturbed clusters tend to exhibit bluer colours compared to those in relaxed systems. This trend is consistent with previous findings indicating that disturbed and dynamically unrelaxed environments can delay environmental quenching, allowing satellites to retain younger stellar populations and bluer colours (e.g. \citealt{Hou2012, Cava2017, Aldas2023, Kelkar2023}). At high-redshift, the distinction becomes more pronounced, likely reflecting a combination of more gas-rich galaxies and the influence of ongoing dynamical interactions that can sustain-or even trigger-star formation.

Comparing the distributions at high- and low-redshift, we can see that in general, as expected by the Butcher-Oelmer effect, the blue galaxy population is more prominent at higher redshifts. Focusing on galaxies located outside $R_{200}$, we observe that, at high-redshift, disturbed clusters are dominated by blue galaxies, whereas in relaxed clusters, the red sequence remains more prominent.

\begin{figure*}
		\centering
		\includegraphics[width=\linewidth, trim={3cm, 2.3cm, 3.5cm, 1cm}, clip]{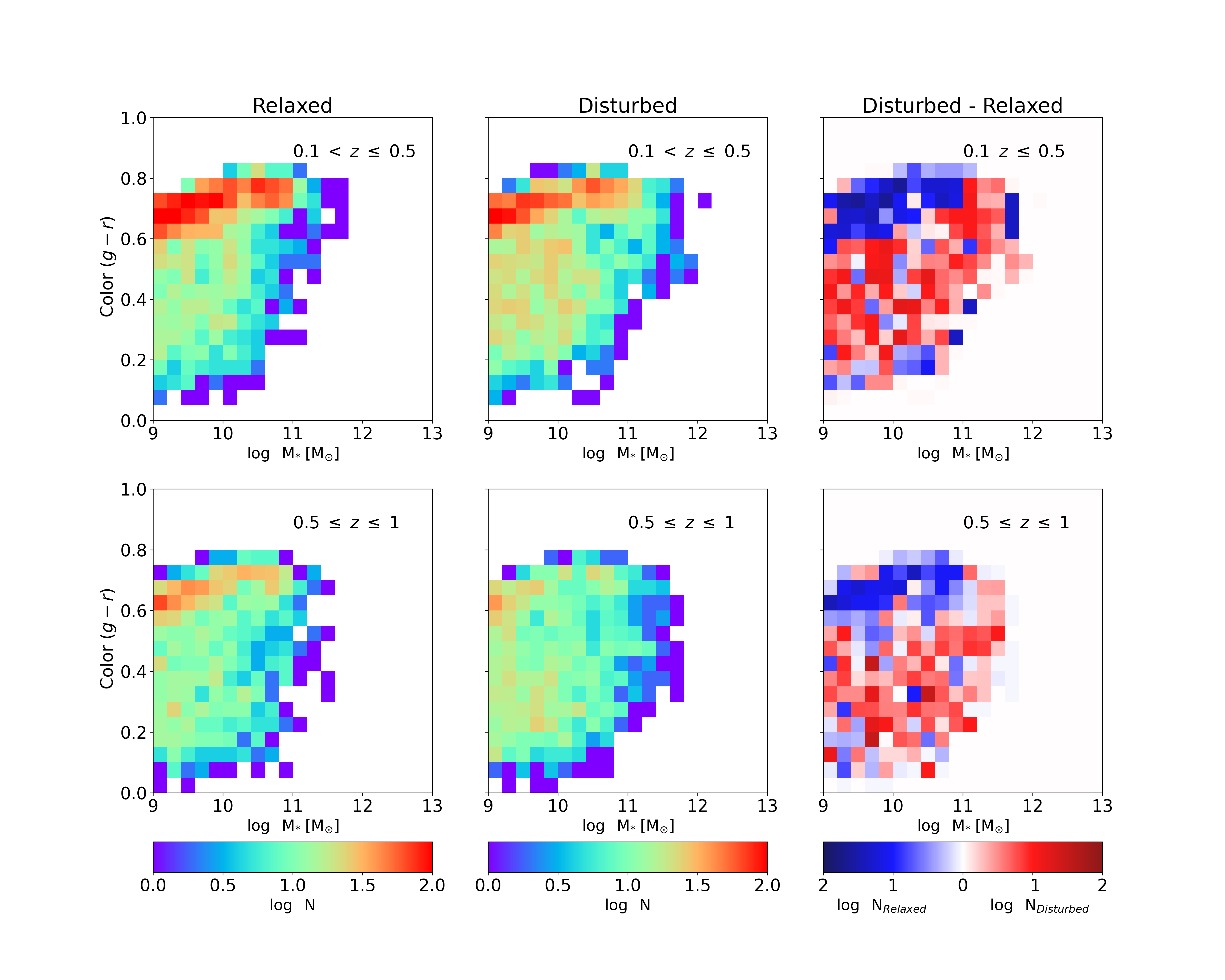}
		\caption{Colour versus stellar mass diagram distribution of galaxies in relaxed in the left panel and disturbed in the middle one, for the low-redshift (top) and high-redshift (bottom) bins. 
  In the left and middle panels we show the satellite galaxy densities distribution for the disturbed and relaxed clusters; whereas in the right panels,  
  we have the difference in the galaxy density between relaxed and disturbed galaxy clusters after normalising. Notice that the sum of the densities in the last panel equals zero. There is an excess of galaxies in the red sequence of the diagram in relaxed clusters compared with the disturbed ones.  }
		\label{Fig:ColorvsSM}
\end{figure*}
 
In the second row of Figures~\ref{Fig:LR_ColorvsSM} and \ref{Fig:HR_ColorvsSM}, we present the distribution of galaxies' sSFR, defined as the SFR divided by its stellar mass.  Note that not all galaxies are shown in this histograms, as a fraction of them are fully quenched. However, all galaxies are included in the normalisation of the distributions. In the low- (high-) redshift bin, $56\%$ ($38\%$) of satellite galaxies in relaxed clusters are devoid of gas, and $61\%$ ($43\%$) exhibit no ongoing star formation. In contrast, in disturbed clusters, only $35\%$ ($21\%$) are gas-exhausted galaxies and $40\%$ ($24\%$) have null star formation. As a result, those plots show just $39\%$ ($57\%$) of satellite galaxies in relaxed clusters and $60\%$ ($76\%$) in disturbed clusters.  We clearly observe that, in disturbed clusters, a significant fraction of star-forming galaxies lies outside $\text{R}_{200}$. Moreover, in these outer regions, disturbed clusters host a larger number of galaxies with sSFR between 10$^{-9.5}$ to  10$^{-10.5}$ [yr$^{-1}$] compared to relaxed clusters. However, for galaxies inside $\text{R}_{200}$, the sSFR distributions of the star-forming population are similar in both relaxed and disturbed clusters, across both redshift bins. Finally, it is worth noting that, as expected, high-redshift clusters host a larger fraction of star-forming galaxies and galaxies that contain gas, compared to their low-redshift counterparts.

Metal elements are synthesised in the stellar interiors and then redistributed into the Interstellar medium through processes such as SN Type I and Type II and AGB winds. Thus, the gas metallicity provides insights into the star formation history in galaxies~\citep{Searle1972, Nagao2006}. The histograms in the bottom panels of Figures~\ref{Fig:LR_ColorvsSM} and (\ref{Fig:HR_ColorvsSM})  show the mean gas iron content distribution, denoted as $\log([Fe/H])$ of galaxies in both relaxed and distrurbed clusters. We see that all distributions range from 0 to 0.4 and peak at similar metallicity values $\log ([Fe/H])\approx 0.2$, meaning around 1.6 times the solar metallicity. Also, the distribution of metallicity in relaxed and disturbed clusters is similar for the inner regions of the clusters in both redshift bins. We note that, although metallicity distributions show that galaxies in disturbed clusters are slightly more metal-poor than those in relaxed clusters-particularly at higher redshift and in the outer regions-this may be due to the continued accretion of less chemically evolved galaxies from the field and delayed chemical enrichment~\citep{Pasquali2012TheSurvey}.

To better visualise the differences in the distribution of galaxies in relaxed and disturbed clusters, we present, in Figure ~\ref{Fig:ColorvsSM},  their density distribution in colour (g-r) versus stellar mass, considering all galaxies inside and outside $\text{R}_{200}$, for the low- (high-) redshift bin in the top (bottom) panels. In the left and middle panels, we show the results for the relaxed and disturbed clusters, respectively. Those  panels are normalised in a way that, at the same redshift bin, clusters in relaxed and disturbed clusters have the same number of galaxies.  The most representative feature in those panels is the RS, which is the zone with the higher density of galaxies around $(g-r)\approx 0.7$, this zone is present in all of the panels but it is much better defined in the low-redshift ones. This zone is populated by red/elliptical galaxies, whose stellar populations are mainly composed of red and old stars. For this reason, they cover a small range in colour but a larger range in magnitudes.  Under the red sequence, we can also identify the blue cloud corresponding to galaxies with recent stellar formation activity.   To highlight the difference between those two distributions, we show the residual map in the rightmost panels after subtracting the normalised number between the galaxy population of relaxed clusters from the disturbed clusters in each bin. These panels clearly shows that the red sequence is significantly more populated in relaxed clusters, while the remaining area, including the blue cloud, is more prominent in disturbed clusters. This means that there is more recent formation activity in the disturbed clusters than in the relaxed ones. This conclusion holds for both high- and low-redshift bins. Although the RS is more clearly defined at low redshift, the residuals exhibit similar patterns across both bins, suggesting that galaxies in relaxed clusters are more quenched than those in disturbed clusters, regardless of redshift.

\begin{figure*}
		\centering
		\includegraphics[width=\linewidth, trim={3cm, 2.3cm, 3.5cm, 2cm}, clip]{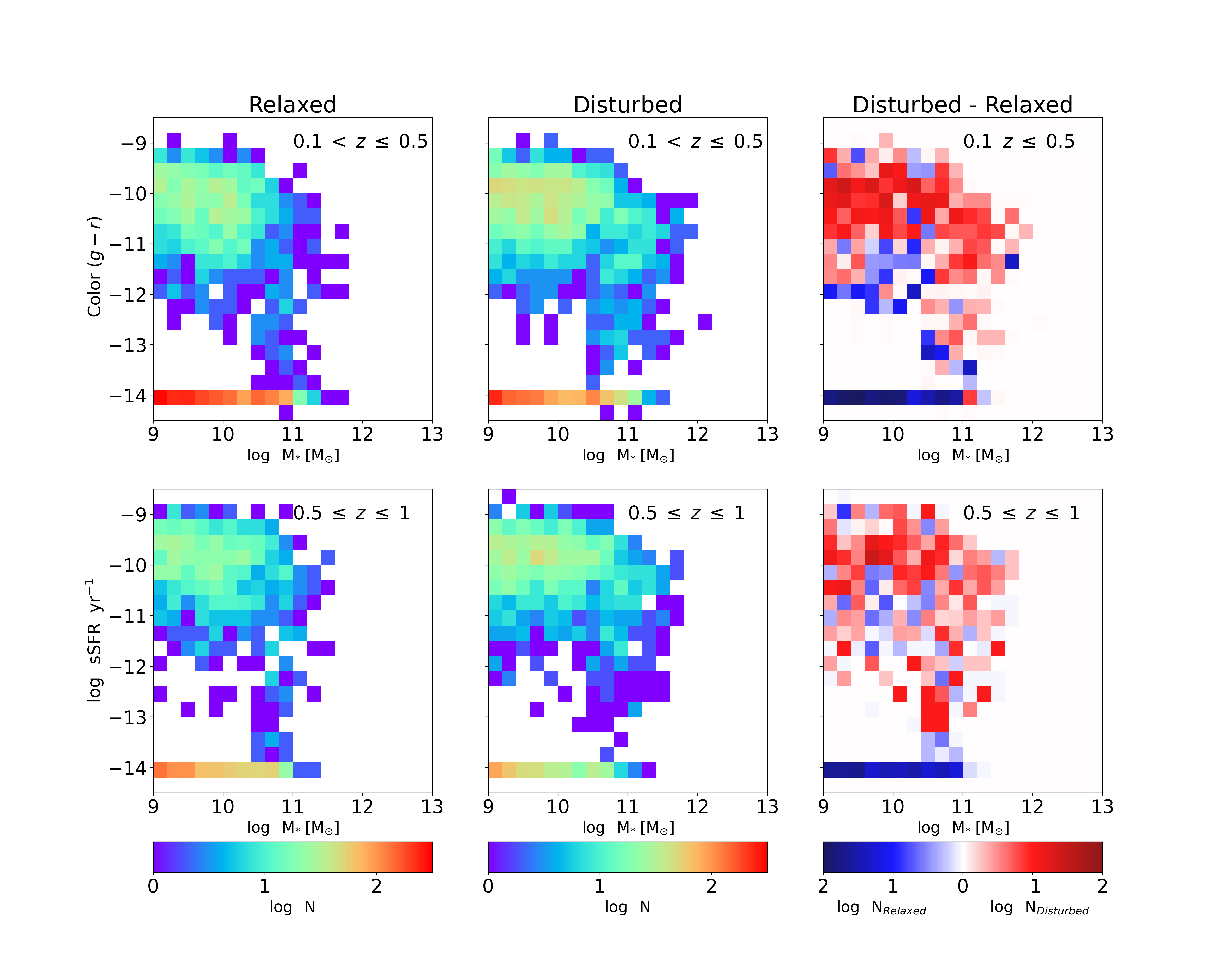}
		\caption{Specific star formation rate in function of stellar mass diagrams for relaxed halos in the left panels, disturbed halos in the centre panels, and the difference between relaxed and disturbed halos right panels, for the low-redshift (top) and high-redshift (bottom) bins. In the last panel, the excess of galaxies bounded to relaxed halos are shown in blue, and the zones of the diagram with an excess of galaxies bounded to disturbed halos are shown in red. There is an excess of quenched galaxies in relaxed clusters compared with the disturbed ones. }
		\label{Fig:sSFRvsSM}
\end{figure*}

Figure~\ref{Fig:sSFRvsSM} presents similar diagrams, but the sSFR versus the stellar mass distribution in the top panels for the low-redshift bin, while in the bottom panels for galaxies in clusters in the high-redshift bin. Galaxies belonging to relaxed and disturbed clusters are shown in the left and middle panels, respectively. Here, we also include galaxies that have null sSFR with a representative value of $10^{-14}$ yr$^{-1}$. In those plots, we can see the star-forming MS around 10$^{-10}$ yr$^{-1}$. This zone is more populated than other parts of the diagram, except for those that are considered to be fully quenched. However, we notice that it is better defined in the disturbed clusters than in the relaxed ones, this applies for both redshift bins. We can also identify galaxies with SFR between $10^{-14}$ yr$^{-1}$ and $10^{-12}$ yr$^{-1}$ transitioning from the star-forming zone to the quenched zone.  In the right panels, we show the differences between the number of galaxies in relaxed clusters and the number of galaxies in disturbed clusters.  We can see that, with respect to the disturbed clusters, the zone of quenched galaxies is more populated by galaxies in relaxed clusters. In contrast, in all other areas, i.e. the star-forming and transition zones, are dominated by the galaxies in disturbed clusters. Again, those results are valid for both low- and high-redshift bins.
The distribution of galaxies in Figures \ref{Fig:ColorvsSM}, and \ref{Fig:sSFRvsSM}, and the fact that we observed a clear difference in the fraction of red and quenched galaxies in relaxed clusters and more blue and star-forming galaxies in disturbed clusters, suggests that relaxed clusters are in a more evolved dynamical state than disturbed clusters.

\section{Quenched fraction}
\label{Sec:Redshift}
Galaxy clusters at low-redshift ($z<0.5$) are generally more relaxed and dominated by early-type, quenched galaxies, whereas high-redshift ($z>0.5$) clusters exhibit more substructure and host a larger fraction of blue, star-forming galaxies \citep{Butcher1984, Postman2005, Poggianti2006, Zenteno2020, Pallero2022}. These trends reflect the progressive assembly of cluster structures and the delayed onset of environmental quenching at earlier epochs \citep{DeLucia2024, Haines2015, Goto2003}. To investigate this reported difference, we analyse  in this section the quenched fraction of satellite galaxies as a function of redshift and examine its  dependence on the dynamical state of their host groups and clusters.

\begin{figure}
    \centering
    \includegraphics[width=\linewidth, trim={0.8cm 2cm 2.4cm 3cm}, clip]{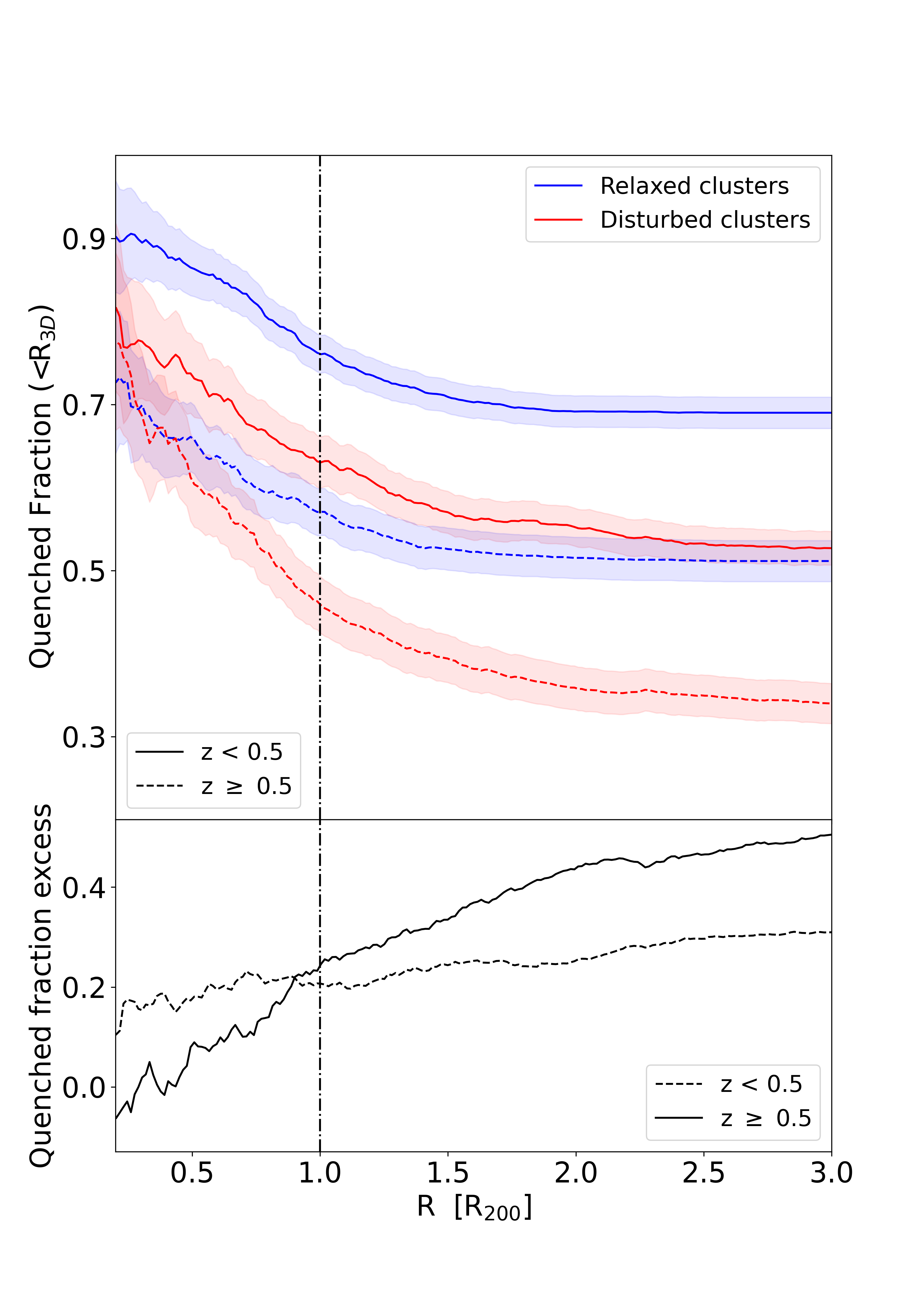}
    \caption{Top panel: Quenched fraction in the function of the 3D radius from the cluster's centre. In solid lines, the quenched fraction of galaxies that inhabit the low-redshift clusters ($z < 0.5$), and in dotted lines, the fraction of quenched galaxies for clusters in high-redshift bins ($z\geq0.5$). Quenched fractions of galaxies in relaxed clusters are presented in blue, and galaxies inside disturbed clusters are presented in red. The shadow areas are the Poisson's noise. Bottom panel: Relaxed quenched fraction excess for the high-redshift bin with a solid line and for the high-redshift bin with a dotted line. }
    \label{Fig:Quenched_fraction}
\end{figure}

The top panel of Figure~\ref{Fig:Quenched_fraction} presents the quenched fraction of satellite galaxies as a function of distance from the cluster centre (in units of $\text{R}_{200}$) for both relaxed and disturbed clusters across different redshift ranges. This distance is the true 3D separation from the cluster centre, ensuring that our analysis accounts for the full spatial distribution of galaxies within the cluster environment. The cluster centre corresponds to the position of the most bound particle within the halo, identified using the Friends-of-Friends (FoF) halo-finder algorithm. The most bound particle coincides with the position of the BCG, as identified by SUBFIND. This particle represents the lowest point in the gravitational potential well and is a robust marker for the cluster’s dynamic centre. In relaxed clusters, the most bound particle is typically well aligned with the centre-of-mass of the intracluster gas. However, in disturbed clusters, ongoing mergers and dynamic interactions can cause significant offsets between these quantities. 

The galaxies were classified as quenched using the definition provided in section~\ref{Section:Quenched_definition}. The fraction of quenched satellite galaxies was computed as the number of passive galaxies divided by the number of all galaxies whose holocentric distances are lower than R. Blue and red lines denote relaxed clusters and disturbed clusters, respectively, and solid lines indicate the results for the low-redshift ($z<0.5$) sub-samples, while dashed lines do for high-redshift ($z\geq0.5$) sub-samples. The shaded regions around each line indicate the uncertainties in the quenched fraction computed under the assumption of Poisson noise. \\

We define the central regions of clusters as the volume within R$_{200}$. This choice is motivated by the fact that R$_{200}$ approximately corresponds to the virialised region of relaxed clusters, where the ICM is in approximate hydrostatic equilibrium. The outer regions of clusters is defined as those with $\text{R}>\text{R}_{200}$, where the influence of infalling structures and accretion processes becomes more significant. We included satellite galaxies up to $3\times \text{R}_{200}$, because even at those large radius, galaxies can be influenced by the cluster environment. In this sense, ~\citet{Sifon2024} found that SFRs in galaxies remain lower than expected for field galaxies even at such large distances. Similarly, ~\citet{Piraino-Cerda2024} demonstrated that galaxies can experience gravitational interactions with clusters beyond $4\times \text{R}_{200}$, identifying clear signatures. The top panel of Figure~\ref{Fig:Quenched_fraction} illustrates that the quenched fraction declines with increasing distance from the cluster centre for both relaxed and disturbed clusters, although relaxed clusters consistently display higher quenched fractions. This trend is more pronounced at low redshifts, indicating that environmental quenching processes are more effective or have had more time to act at these redshifts. We refer to environmental quenching as the suppression of star formation caused by process acting within the cluster environment. Key mechanisms include RPS, which removes the cold gas reservoirs of infalling galaxies, tidal interactions, which disturb their morphology and dynamics, and strangulation, which halts star formation by cutting off the supply of fresh gas. In our sample, in the low-redshift bin, the fraction of quenched galaxies in the central region of the cluster is approximately 90 \%, gradually decreasing to 70\% considering all galaxies within a radial distance of $3 \times \text{R}_{200}$. Meanwhile, for relaxed clusters, the fraction of quenched galaxies in the central regions is around 80 \%, diminishing to 52 \%, in the cluster outskirts. On the other hand, in the high-redshift bin, both relaxed and disturbed clusters manifest a quenched galaxy fraction of around 75\% in their central regions, which subsequently decreases to 54\%, taking into account satellite galaxies up $3\times \text{R}_{200}$ of the relaxed clusters and 36\% for disturbed clusters. The higher quenched fractions observed in the central regions of both relaxed and disturbed clusters suggest the presence of strong environmental effects such as RPS and tidal interactions close to the cluster core~\citep{Boselli2006}. This trend is consistent with the observational study of~\citep{Einasto2018}, which found that galaxy population in the supercluster A2142 near the cluster centre tend to be older than galaxies in the outskirts, most star-forming galaxies are located at distances larger than 1.8 $\text{h}^{-1} \text{Mpc}$.

To conduct a more robust comparison between the quenched fraction in relaxed and disturbed clusters, and noting that the quenched fraction consistently exhibits higher values in relaxed clusters compared to disturbed ones, we introduce the concept of the relaxed quenched fraction excess  $(f_{Q \text{excess}}^{\text{rel}})$ following \cite{Wetzel2013}, defined as: 
\[
    f_{Q \text{excess}}^{\text{rel}}= \frac{f_Q^{\text{rel}}-f_Q^{\text{dist}}}{f_Q^{\text{dist}}}.
\]

The relaxed quenched fraction excess, presented in the bottom panel of Figure~\ref{Fig:Quenched_fraction},  quantifies the excess (or deficit) of quenched galaxies in relaxed clusters with respect to those typically found in their disturbed counterparts. The bottom panel shows that quenched fraction excess for relaxed clusters increases with distance from the cluster centre, especially at high-redshifts. This suggests that, especially at $z\geq 0.5$,  quenching in relaxed systems is more effective even in the outskirts of clusters, potentially impacting satellite galaxies that are on infalling trajectories or those that have been ejected but remain within the cluster’s gravitational influence \citep{Bahe2013}. Although the quenching effects are much stronger around $\text{R}_{200}$, in this sense, ~\citet{Pallero2022} show that ram pressure rapidly dominates the restoring force after galaxies cross the virial radius, leading to rapid gas depletion and quenching. \citet{Wetzel2014} show that the environmental quench could extend up to $2.5\times \text{R}_{200}$ and can be explained by satellites ejected by the cluster.  In contrast, at lower redshifts, the excess remains relatively stable, reflecting a sustained quenching effect throughout the cluster's extent. This stability aligns with~\citet{Pallero2022}, who found that most of satellites are quenched during their initial infall, indicating that initial interactions with the cluster environment are crucial for quenching.

\begin{figure}
    \centering
    \includegraphics[width=\linewidth, trim={0.97cm 0.45cm 1cm 1cm}, clip]{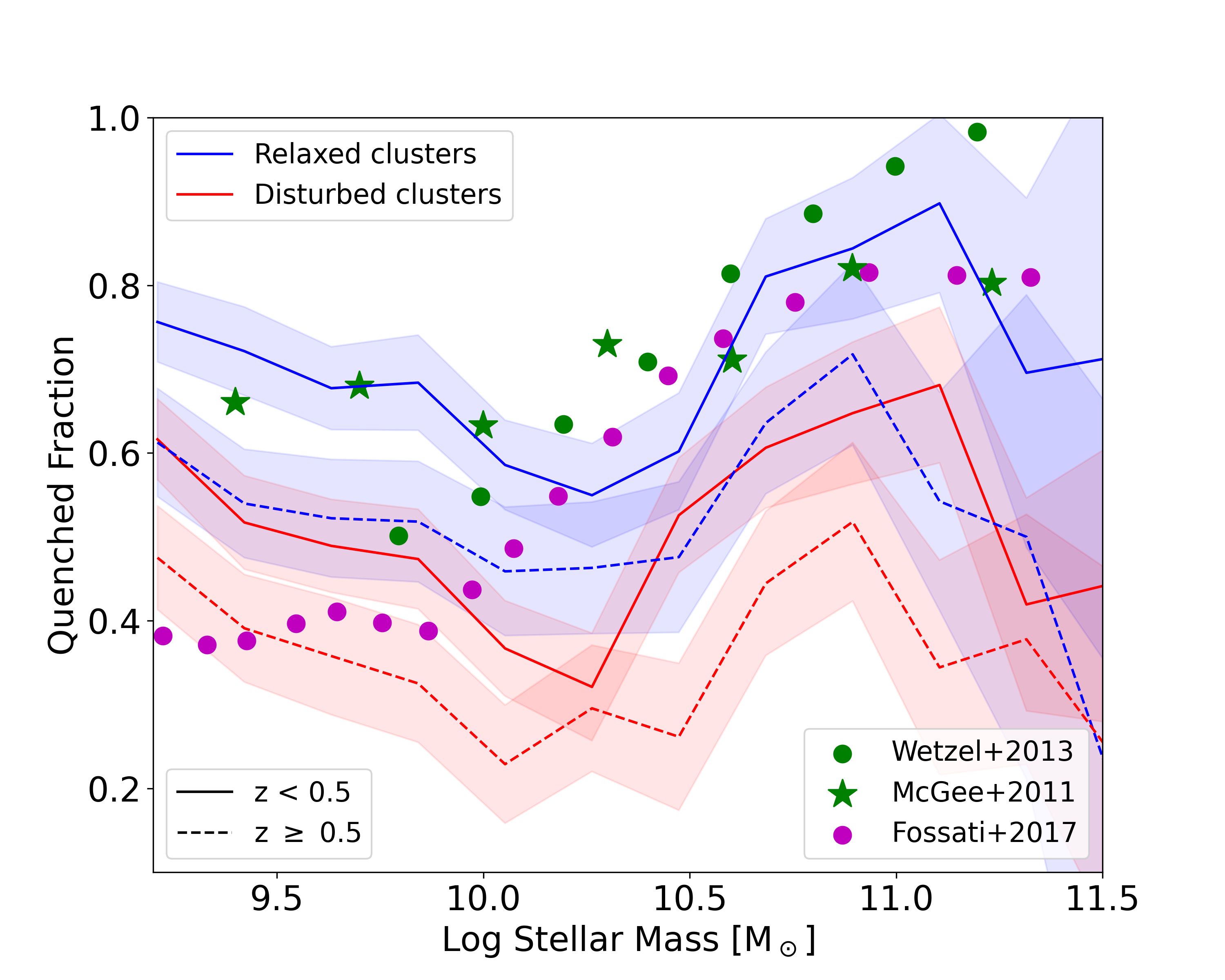}
    \caption{Quenched fraction of galaxies as a function of stellar mass. In blue, we present the galaxies bounded to relaxed halos, and in red, we present the galaxies bounded to disturbed halos. We separated our sample between low ($z < 0.5$) redshift sample shown in solid lines and high-redshift ($z \geq 0.5$) bins shown in dashed lines. The shadow areas represent the uncertainty computed using Poisson's noise. Comparing the sub-samples at the same redshift, the relaxed groups and clusters always have more quenched galaxies that disturbed ones.  }
    \label{fig:QFStellarmass}
\end{figure}

In Figure \ref{fig:QFStellarmass}, we show the quenched fraction of satellite galaxies as a function of galaxy stellar mass. At low redshifts, the quenched fraction in relaxed clusters (solid blue line) shows a slight decrease for masses between $10^{9}$ and $10^{10} \text{M}_{\odot}$, followed by a significant increase with stellar mass, peaking around $\text{M}_{*}\sim 10^{11} \text{M}_{\odot}$. The quenched fraction for disturbed clusters at low redshifts (solid red line) follows a similar pattern, but remains consistently lower than in relaxed clusters. This difference is present at low- and high-redshift bins, suggesting that the more stable environments (relaxed clusters) facilitate more effective quenching.
We compared our simulated quenched fraction with various observational datasets. For the low-redshift bin, the data from \citet{Wetzel2013} (green circles) correspond to satellite galaxies bound to low-redshift clusters (up to $z=0.06$), based on the NYU Value-Added Galaxy Catalogue from the Sloan Digital Sky Survey (SDSS). These observational results align well with the quenched fraction in relaxed clusters, especially at higher stellar masses, but show discrepancies in the low mass regime. As demonstrated by \citet{Donnari2021}, these differences can be attributed to the sample selection. In this context, they illustrated that Wetzel's data could be better replicated by including satellite galaxies bound to lower mass groups ($\text{M}_{200}>10^{11} \text{M}_{\odot}$).

Furthermore, comparing with \citet{McGee2011} (green stars), using SDSS data at $z\sim 0.08$, it shows that our simulated quenched fractions at low-redshift relaxed clusters follow a similar trend. Although their work does not specify any mass cutoff for the galaxy groups, they infer that they are including galaxies belonging to groups with masses $\text{M}>10^{12.75} \text{M}_{\odot}$. Therefore, this closer agreement between our simulated quenched fraction and McGuee's observational data is likely due to the better match between the sample selection criteria.

The data from \citet{Fossati2017} (purple circles) correspond to high-redshift observations. Specifically, we are comparing the passive galaxy fraction for satellite galaxies with host masses $\text{M}>10^{13} \text{M}_{\odot}$ in the redshift range of $0.5<z<0.8$. Our simulations reasonably match the observed data, particularly at lower stellar masses. However, the simulations underestimate the quenched fraction in the high mass range. Nevertheless, it is worth noting that the simulation can reproduce a lower quenched fraction in the high-redshift bin compared to the low-redshift bin at low galaxy mass regimen, as obtained from observational data.

Figure~\ref{fig:QFStellarmass} illustrates a decrease in the quenched fraction at stellar masses exceeding $10^{11},\text{M}_{\odot}$. This deviation arises from numerical noise caused by low statistical significance within this mass regime. As demonstrated by \citet{Donnari2021}, employing a substantially larger galaxy sample drawn from the TNG300 simulation and using comparable halo selection and quenched galaxy criteria, this deviation disappears.
  
The higher fraction of quenched galaxies in relaxed clusters compared to disturbed ones can also be explained, at least in part, by dynamical interactions that facilitate the infall of star-forming galaxies from the surrounding cosmic web into disturbed systems. This continuous accretion can replenish the population of star-forming galaxies, thereby reducing the overall quenched fraction in these clusters. Moreover, quenching mechanisms operate in different regimes of redshift and stellar mass, as highlighted by \citet{Peng2010}. At lower redshifts and higher stellar masses, mass quenching—regulated by internal processes such as AGN feedback and supernova outflows—tends to dominate. In contrast, at higher redshifts and lower stellar masses, environmental quenching—driven by ram-pressure stripping, strangulation, and tidal interactions—becomes more significant. Given that disturbed clusters are more likely to be in an active phase of assembly, with ongoing mergers and recent accretion of satellite galaxies, their overall quenching efficiency may be systematically lower than that of relaxed systems, where galaxies have had more time to be processed by environmental effects.

In summary, this section's results underscore the influence of the cluster's dynamical state on satellite galaxy quenching. Relaxed clusters exhibit higher quenched fractions at both redshifts, suggesting that stable environments enhance the efficiency of quenching mechanisms. The difference between relaxed and disturbed clusters is more pronounced at low redshifts, where mature clusters have more time to quench their satellite galaxies effectively.

\section{Gas content}
\label{Sec:Gas}
Here, we analyse the gas content of satellite galaxies within both relaxed and disturbed clusters and groups in the IllustrisTNG simulations. Quantifying gas across various phases yields valuable insights into the physical processes driving galaxy evolution and may offer crucial indicators of a cluster's dynamical state \citep{Pillepich2017model, Nelson2019}. The availability of cold gas is crucial for star formation, directly influencing a galaxy's evolutionary trajectory. Cold gas serves as the fuel for star formation; its depletion leads a decline in star formation activity, and ultimately, to quenching. Conversely, the inflow of fresh cold gas replenishes the reservoir, sustaining-or even reigniting-star formation. Several key processes regulate this balance: gas accretion from the surrounding cosmic environment provides the raw material~\citep{Nipoti2009}, while star formation itself consumes this gas~\citep{Hoyle1953}. Feedback mechanisms, such as those driven by supernovae and AGN, can heat and expel cold gas, thereby regulating star formation~\citep{Muller-Sanchez2018}. Finally, environmental effects, particularly within galaxy clusters, can strip gas from their galaxies through processes like ram pressure and tidal stripping.
As previously discussed, stellar populations born at different times will have different characteristics depending on the properties of the intra-galaxy gas. For example, gas clouds are progressively enriched with chemical elements, thanks to processes such as SN Type I and II, and winds from AGB stars. As a result, galaxies with ongoing start-formation activity will have not only bluer stellar populations, but also different chemical compositions. 

To characterise the gas available in galaxies bound to relaxed and disturbed clusters, we classify it into diffuse, condensed, and hot phases based on density and temperature. The temperature of the gas is calculated for each gas cell using the internal energy $\mu$ and the Electron Abundance $(x_e= n_e/n_H)$, provided in the simulation outputs with the following equation 
\[
    T=(\gamma -1) u/k_B \times \mu.
\]
Here $k_B $ is the Boltzmann constant, $\gamma = 5/3$,  $X_H=0.76$, $m_p$ is the proton mass, and  $\mu$ the mean molecular weight, defined as:
\[
    \mu= \dfrac{4}{1+3X_H+4X_H x_e}\times m_p.
\]

The classification of the different gas phases applied here is based on the boundaries proposed by \citet{Torrey2012} and \citet{Torrey2019TheIllustrisTNG}. Gas is considered hot if $\log(T/10^{6} K) > 0.25 \log (n_H /405 cm^{-1})$, where $n_H$ is the gas density expressed in units of atoms of Hidrogen per unit of cubic volume. Gas not meeting this condition is further categorised as diffuse if its density is $n_H<0.13 cm^{-1}$, and as condensed if its density is greater than $n_H > 0.13 cm^{-1}$.

\begin{figure}
		\centering
		\includegraphics[width=\linewidth, trim={1.5cm, 1.9cm, 2.0cm, 2.5cm}, clip]{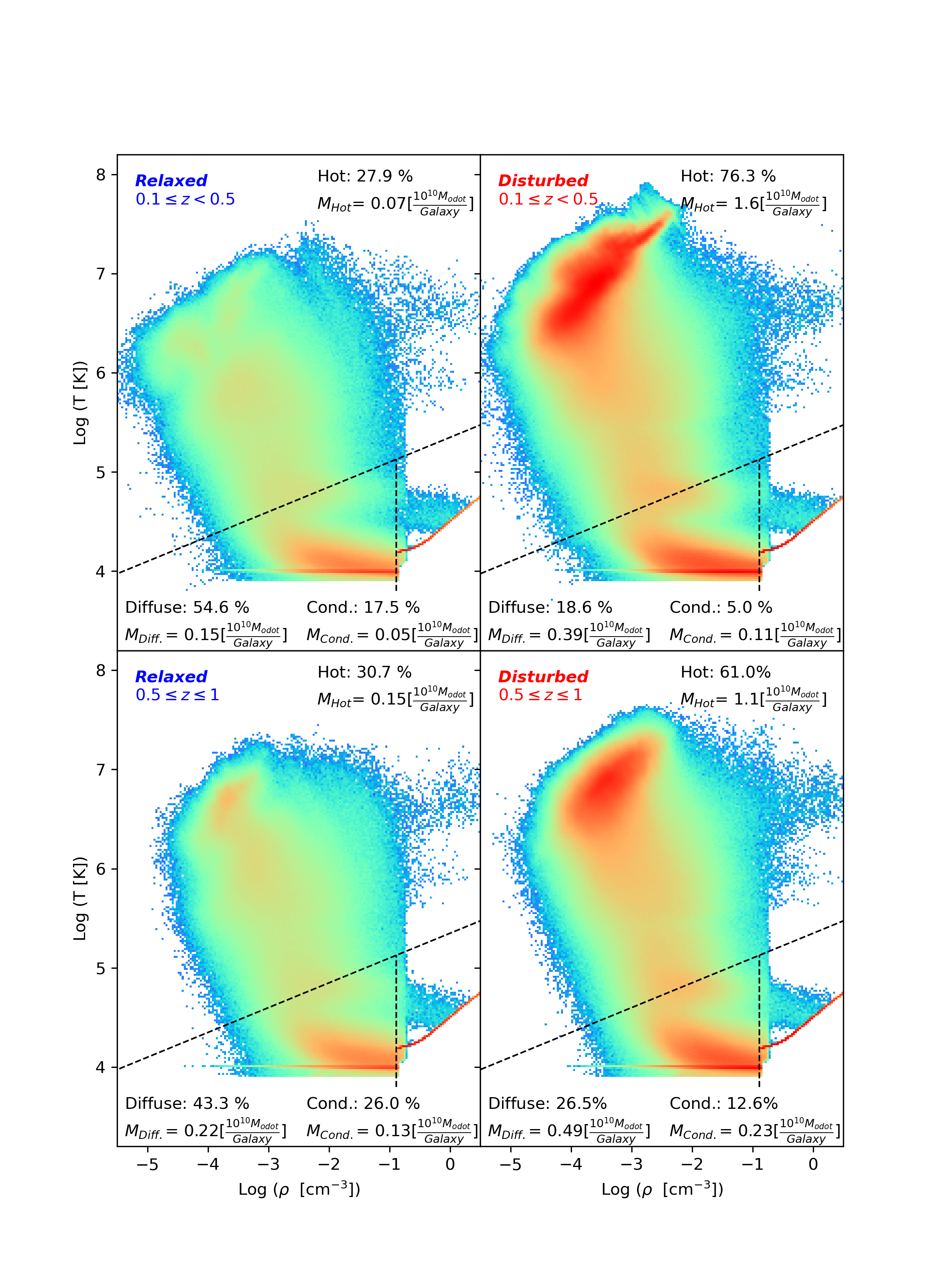}
		\caption{Gas content of the satellite galaxies in relaxed (left panels) and disturbed clusters (right panels), for low- (upper panels), and high-redshift (bottom panels). Each panel shows the gas content for each gas phase: hot, diffuse, and condensed. On average, galaxies in disturbed clusters have more gas than galaxies in relaxed ones. We also can see that galaxies in disturbed clusters have more hot gas due to the inclusion of the secondary BCG and the gas heat suffered during the interaction.}
		\label{Fig:GasTemp}
\end{figure}

In Figure~\ref{Fig:GasTemp}, the four panels show the distribution of gas related to galaxy clusters in different categories based on their dynamical state (relaxed or disturbed) and redshift (low or high). Those diagrams were obtained by adding together all of the gas particles bound to the cluster galaxies, excluding the gas of the BCG. The x axis represents the logarithm of gas density ($n_H$ in cm$^{-3}$), and the y axis displays the logarithm of gas temperature (T in K). Each panel is colour-coded to represent the amount of gas, with blue indicating lower gas quantities and red indicating higher quantities. Additionally, each diagram includes the percentages of gas in the different phases relative to their total and the average amount of gas per galaxy in the corresponding phase per galaxy satellite. The dashed lines demarcate the boundaries between the different gas phases. 

Figure~\ref{Fig:GasTemp} allows us to understand the relationship between the amount of gas available and the dynamical state of the clusters. Satellite galaxies in groups and clusters contain significant amounts of gas across all gas phases. When comparing satellite galaxies in relaxed and disturbed clusters, it is evident that disturbed clusters contain more gas in the hot phase in both low- and high-redshift bins compared to the relaxed ones. The difference in the hot gas fraction is quite substantial, with 28\% for relaxed clusters and 76\% for disturbed clusters in the low-redshift bin. In the high-redshift bin, the hot gas fraction is 30\% for relaxed clusters and 61\% for disturbed clusters. It is important to note that not only is the fraction of hot gas greater in disturbed halos, but more importantly, the average mass per galaxy is much higher.

We now look at the diffuse gas phase, corresponding to cold gas cells with a density lower than that required to form stars. In this phase, disturbed clusters have a significantly higher average amount of gas per galaxy than relaxed clusters despite having a lower diffuse gas fraction. For instance, in the low-redshift bin, there is 2.6 times more diffuse gas in galaxies belonging to disturbed clusters compared to relaxed ones. This difference is also observed in the high-redshift bin, where disturbed cluster galaxies have 2.2 times more diffuse gas available than relaxed ones. It is crucial to quantify the availability of gas in this phase because, during cluster interactions, the gas can be compressed in the front half of galaxies due to ram pressure~\citep{Schaye2015,Troncoso-Iribarren2020}. This can lead to a burst of star formation and gas transfer to the trailing half of galaxies. 
Using EAGLE simulations, \citet{Troncoso-Iribarren2020} concluded that this effect is more important for galaxies with stellar masses lower than $10^{10.5}$ M$_{\odot}$ located near $\text{R}_{200}$. This enhancement in star formation in infalling galaxies was observationally confirmed by~\citet{Roberts2022} studying the cluster IC3949 using data from the ALMA MaNGA survey. Galaxies with stellar masses larger than this mass range tend to arrive in their final cluster already preprocessed and quenched by their AGN. Additionally, these galaxies may have very little gas in the diffuse phase, as most of it is heated by the AGN. Consequently, gas compression does not play a significant role in this mass range.

The condensed gas phase is the one that has the pressure and density to begin forming stars. This gas phase distribution has a characteristic thin shape in the density-temperature phase diagram, governed by the \citet{Springel2003} state equation, as shown on the right side of the plots in Figure~\ref{Fig:GasTemp}. The amount of available condensed gas per galaxy in disturbed clusters is around $1.5 $ times the amount of gas available in relaxed ones.  As a result, on average, the SFR is higher in disturbed galaxy clusters than in relaxed counterparts.  This ongoing star formation can explain the presence of a more diverse galaxy population and the excess of blue galaxies found in disturbed clusters compared to the relaxed ones presented in Section~\ref{Sec:Differences}. This difference is present in both redshift bins. However, it is more pronounced for galaxy clusters in the high-redshift bin, suggesting that the excess of blue and young galaxy population is likely more significant in the case of clusters between $0.5<z<1$. This result agrees with \citet{Aldas2023}, where they observed that the galaxies are bluer in the disturbed galaxies for this redshift range. 

The differences in the hot gas fraction found between the relaxed and disturbed clusters presented in Figure~\ref{Fig:GasTemp} can be explained by either a more massive population of galaxies in the disturbed cluster (containing a larger amount of hot gas), as described by \citet{Torrey2019TheIllustrisTNG}, or by gas shock heating mechanism during cluster interactions, injecting energy into the gas reservoir. To understand the cause of this difference and to make a more fair comparison between the relaxed and disturbed clusters, in Figure \ref{Fig:GasTempWOBCG}, we present plots analogous to those in Figure \ref{Fig:GasTemp} 
but now also excluding the secondary BCG. 
A primary distinction evident upon a comparison of the two figures lies in the hot gas phase, where the large amount of hot gas with high temperature and low density, observed in Figure~\ref{Fig:GasTemp} becomes less pronounced when excluding the BCG from the secondary galaxy cluster. Consequently, the hot gas fraction in disturbed clusters decreases from 70\% to 35\% for low-redshift disturbed clusters and from 60\% to 30\% for high-redshift disturbed clusters. Despite this, the fraction of hot gas in disturbed clusters remains larger than that measured in their relaxed counterparts. This result indicates that most of the hot gas excess is given the presence of the secondary BCG which in the case that the disturbed cluster corresponds to a merger, this will corresponds to the BCG of the infalling substructure.

\begin{figure}
\includegraphics[width=\linewidth, trim={0cm, 2.5cm, 1cm, 4cm}, clip]{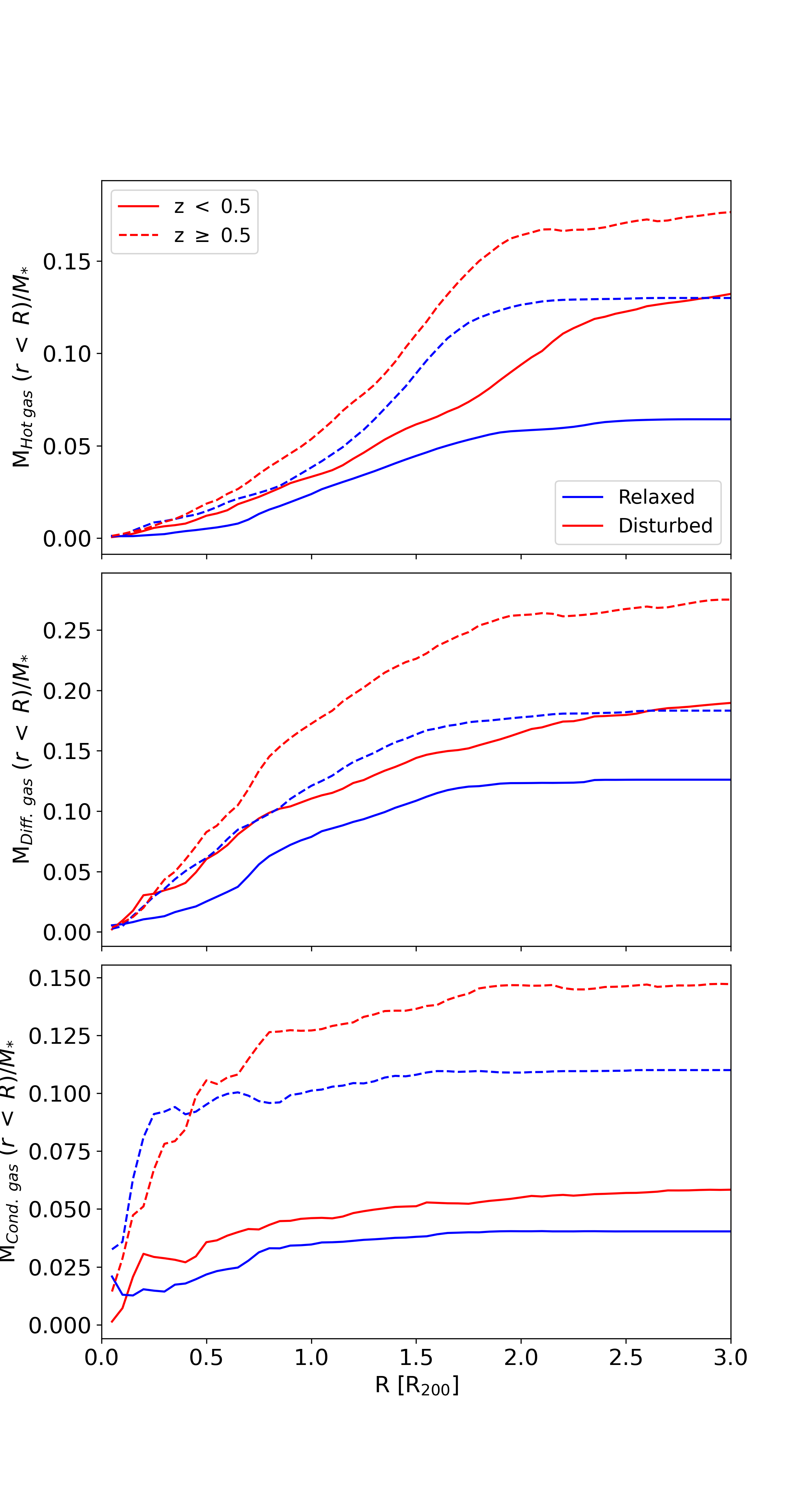}
\caption{Gas fraction as a function of the radius. In the top panel, we have the hot gas as a function of the radius. In the middle panel, we present the warm gas as a function of the radius, and in the last panel, we present the condensed gas, which represents the star-forming gas. }
\label{Fig:Gas_radius}
\end{figure}

Finally, we examine the gas-to-stellar mass ratio, hereafter referred to as the gas ratio. In this analysis, we include gas cells bound to all satellite galaxies, excluding the BCG in relaxed clusters and the main and secondary BCGs in disturbed ones. The gas ratio parameter is defined as the total mass of all gas cells enclosed within a spherical region of radius $\text{R}$, centred on the position of the most bound particle, divided by the total stellar mass within the same volume. It is calculated over the entire enclosed volume rather than in concentric shells, ensuring that it captures the integrated properties of the gas and stellar components within the specified radius. This approach enables a direct comparison of gas content across different clusters while minimising potential biases introduced by localised variations in density. The gas ratio provides insights into the baryon content of the satellite galaxies bounded to the clusters in different dynamical states. This ratio is also closely linked to ram pressure, the predominant mechanism of gas stripping for galaxies in a dense environment like clusters~\citep{Jung2018}.  The results are presented in Figure~\ref{Fig:Gas_radius}, where we show the mean gas ratio for different gas phases: hot gas in the top panel,  diffuse gas in the middle panel, and condensed gas in the bottom panel, all as a function of the cluster centric distance.  Figure~\ref{Fig:Gas_radius} shows that, in the hot gas phase (top panel), satellite galaxies in disturbed clusters retain more hot gas than their counterparts in relaxed clusters, particularly beyond $\text{R} > \text{R}_{200}$ . This excess is more pronounced at high-redshift, likely reflecting shock heating and gas compression within the galaxies themselves as a result of recent cluster assembly activity ~\citep{Molnar2016, Troncoso-Iribarren2020}. The middle panel shows the diffuse gas mass ratio, which is also primarily found in galaxies located in the outer parts of the clusters but closer to the cluster's centre than the hot gas. This panel also indicates that galaxies in relaxed clusters have a larger diffuse gas ratio in disturbed clusters compared to relaxed ones. Lastly, the bottom panel shows the condensed gas distribution, predominantly located in the galaxies in the central regions of clusters. For the high-redshift bin, the condensed gas ratio rises sharply at small radii and then stabilises, whereas in the low-redshift bin, the ratio is relatively flat and minimal. These findings are consistent with the notion that, at lower redshifts, most galaxies are already quenched by the time they are incorporated into their final host clusters~\citep{Pallero2022}.

Our results are consistent with those of~\citep{Stroe2021}, who found that merging clusters host a higher fraction of star-forming galaxies and identified a distinct population of blue, continuum-faint galaxies with high H$\alpha$ equivalent widths, likely undergoing merger-induced starbursts. They also reported a more uniform distribution of AGN and star-forming galaxies in merging clusters compared to relaxed ones, where AGN activity peaks in the outskirts. These observational trends support our  findings that disturbed clusters contain more gas across all thermodynamic phases—hot, diffuse, and condensed—thereby enabling enhanced star formation and AGN activity during mergers.

It is worth noting that by including the secondary BCG in disturbed clusters, the results for the condensed and diffuse gas ratio  are not affected. However, a significant increase in the hot gas ratio is observed, as illustrated in Figure~\ref{Fig:GasTemp}. This indicates that the secondary BCG is primarily composed of hot gas while containing minimal amounts of diffuse and condensed gas. These results support the idea that the populations of galaxies can provide relevant information about the cluster's dynamical state, especially when the two dominant galaxies of the disturbed system are removed from the analysis.

\section{Star formation burst and post quenching.}
\label{Sec:Burst}

\begin{figure*}
    \begin{subfigure}{0.49\linewidth}
        \centering
        \includegraphics[width=\linewidth, trim={0.5cm, 0.5cm, 0.5cm, 0.5cm}, clip]{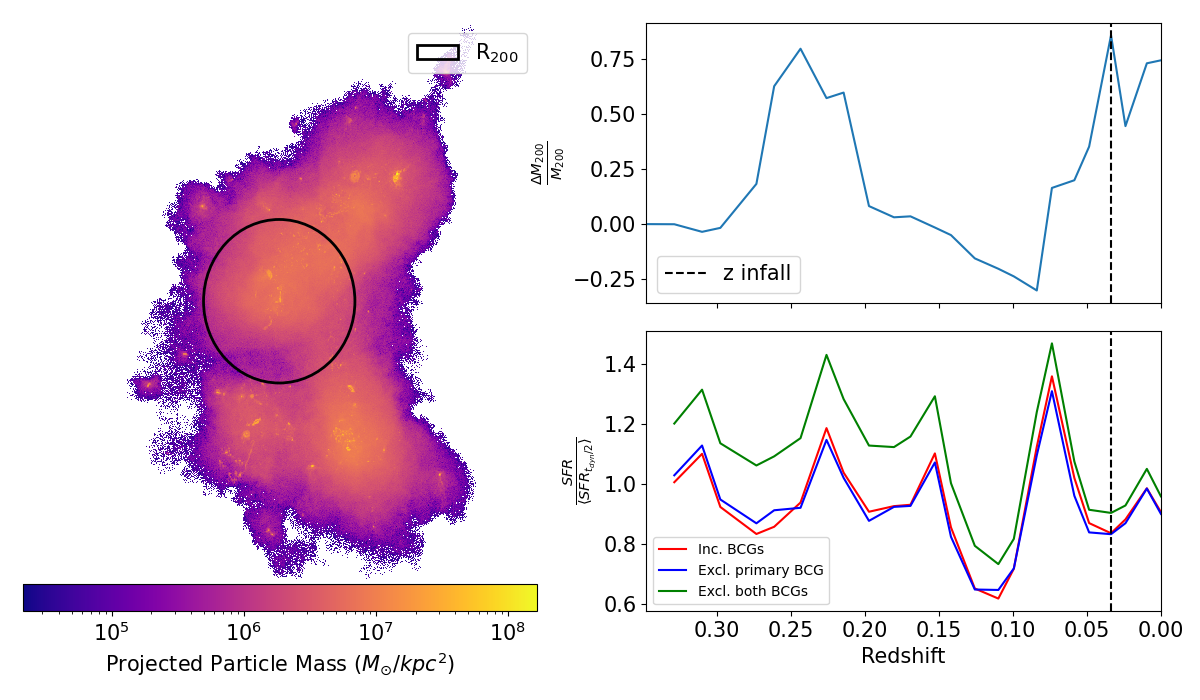}
        \phantomsubcaption
        \label{fig:a}
    \end{subfigure}
\hfill
    \begin{subfigure}{.49\linewidth}
        \centering
        \includegraphics[width=\linewidth, trim={0.5cm, 0.5cm, 0.5cm, 0.5cm}, clip]{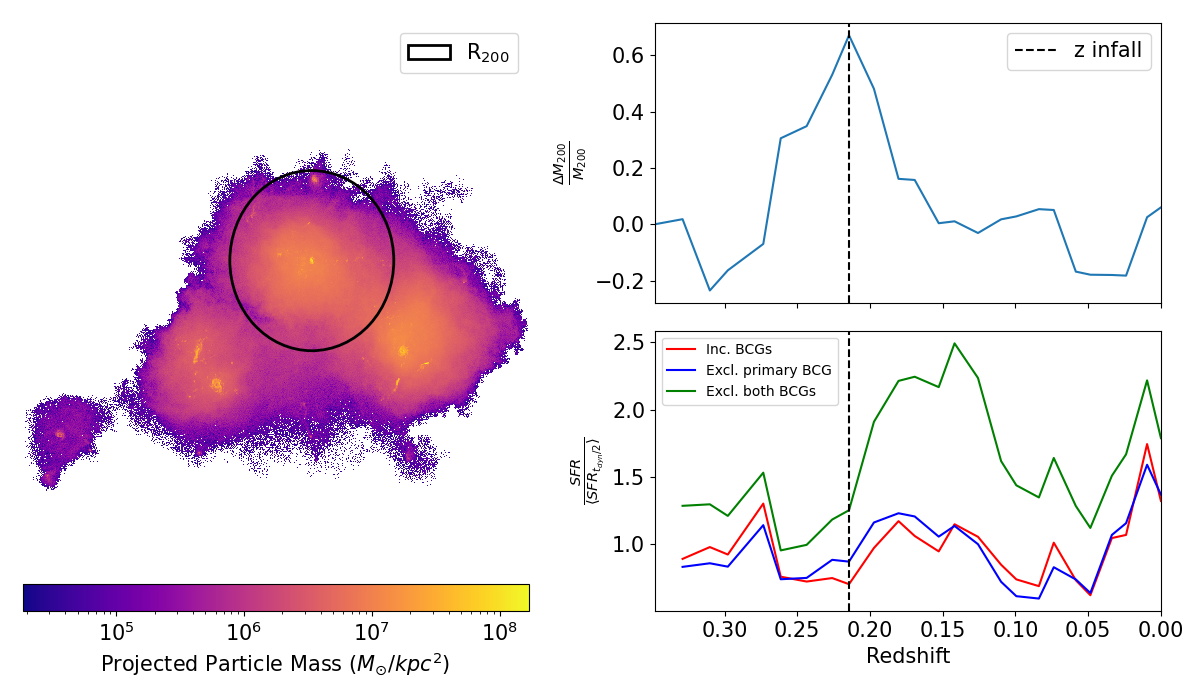}
        \phantomsubcaption
        \label{fig:b}
    \end{subfigure}
\\
    \begin{subfigure}{.49\linewidth}
        \centering
        \includegraphics[width=\linewidth, trim={0.5cm, 0.5cm, 0.5cm, 0.5cm}, clip]{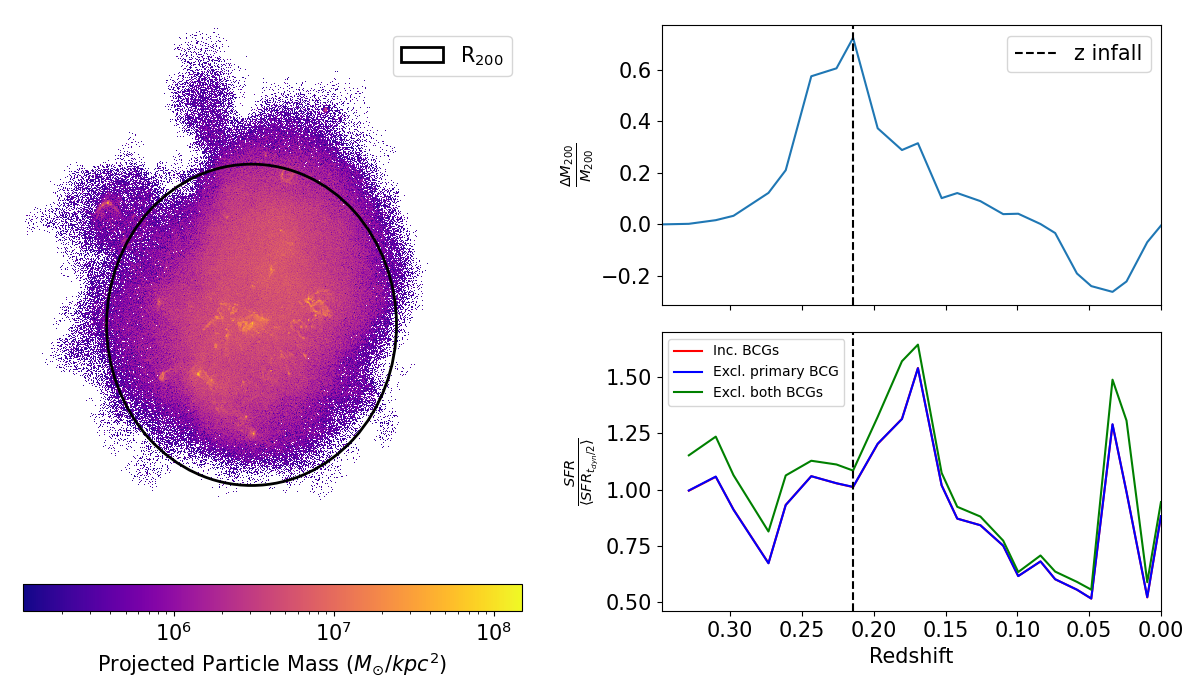}
        \phantomsubcaption
        \label{fig:c}
    \end{subfigure}
\hfill
    \begin{subfigure}{.49\linewidth}
        \centering
        \includegraphics[width=\linewidth, trim={0.5cm, 0.5cm, 0.5cm, 0.5cm}, clip]{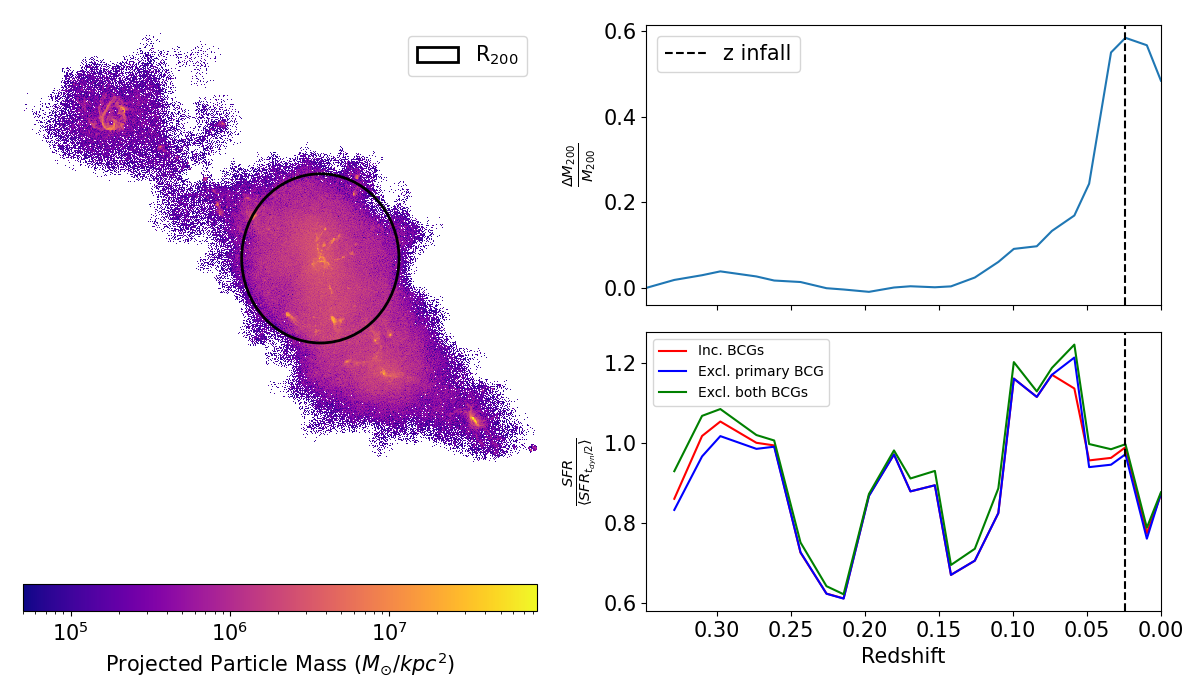}
        \phantomsubcaption
        \label{fig:img}
    \end{subfigure}
\caption{Four selected disturbed galaxy clusters at redshift $z=0.1$; the top ones are clusters in the early interacting stages, while the bottom panels show clusters in more advanced interacting stages where the secondary structure has already crossed $\text{R}_{200}$ of the main one.  On the left side, the projected gas mass distribution for redshift $z=0.1$. On the right side, we have the temporal evolution of total virial mass M$_{200}$ and the sSFR of their satellite galaxies. The clusters exhibit a short boost followed by a decrease in the sSFR, indicating that the sSFR's dynamical state may be influenced by the cluster's dynamical state. }
\label{Fig:evolution}
\end{figure*}

To investigate the origin of the differences in galaxy populations described in Section~\ref{Sec:Differences}, in particular the higher fraction of blue galaxies in disturbed clusters relative to relaxed clusters, this section examines the star formation history and the evolution of the virial mass $\text{M}_{200}$ for clusters classified as disturbed at redshift $z=0.1$, because those provide insights into the later stages of galaxy and cluster evolution as they are closer to their present-day configurations.

Unlike previous sections, where we adopted a general criterion to define unrelaxed clusters, here we focus specifically on disturbed clusters undergoing a merger events. Following the approach of \citet{Contreras-Santos2022} for characterising cluster mergers, we compute the variation in virial mass, $\text{M}_{200}$, for our sample of disturbed clusters over a defined time interval. In this context, we define a merger event as one in which the mass increase satisfies the following condition, 
\[
\frac{\Delta M}{M} = \frac{M_f - M_i}{M_i} \geq 0.5,
\]
within half of the dynamical time. The dynamical time, which depends only on redshift and is independent of cluster mass, was computed following~\citep{Contreras-Santos2022}:
\begin{equation}
    t_d = \sqrt{\frac{3}{4\pi} \frac{1}{200G\rho_c}}.
\end{equation}

To identify potential starburst episodes, we adopt the birthrate parameter, $b$, commonly used in previous studies~\citep{Kohno2005, McQuinn2015}, and defined as:
\[
b = \frac{\mathrm{SFR}}{\langle \mathrm{SFR}_{t_{d}/2} \rangle},
\]
which compares the current SFR to the average rate over the preceding half of the dynamical time.

In Figure~\ref{Fig:evolution}, we examine the evolution of four clusters as a function of time. On the left side of each panel, we show the projected gas mass distribution at $z=0.1$, with a solid line circle indicating the $\text{R}_{200}$ of the main cluster.  It is evident that all four clusters are disturbed systems: their gas morphology is irregular and clumpy, with two or more structures interacting. Some gas clouds appear elongated or exhibit comet-like shapes with tails pointing away from the dense central regions, highlighting the effects of RPS. The top right plot of each panel displays the time evolution of the cluster's $\frac{\Delta \text{M}_{200}}{\text{M}_{200}}$ parameter for the main structure. The vertical dotted lines in those panels mark the redshift at which the parameter reaches its maximum value, which we define as the infall time.  As expected from hierarchical structure formation, the virial mass increases over the time as smaller structures merge to form larger ones. Pronounced increases in the mass parameter over short timescales reflect merger events. The bottom right side of each panel shows the integrated birthrate parameter for cluster galaxies as a function of redshift. This parameter is computed considering both BCGs (red), excluding the primary BCG ( blue), and excluding both BCGs (green). In all cases, peaks in the birthrate parameter followed by rapid declines are observed and coincide with peaks in mass accretion. This correlation suggests that starburst episodes are linked to the cluster merging process.  \\
\begin{figure}
		\centering
		\includegraphics[width=\linewidth, trim={0cm, 0cm, 1cm, 1cm}, clip]{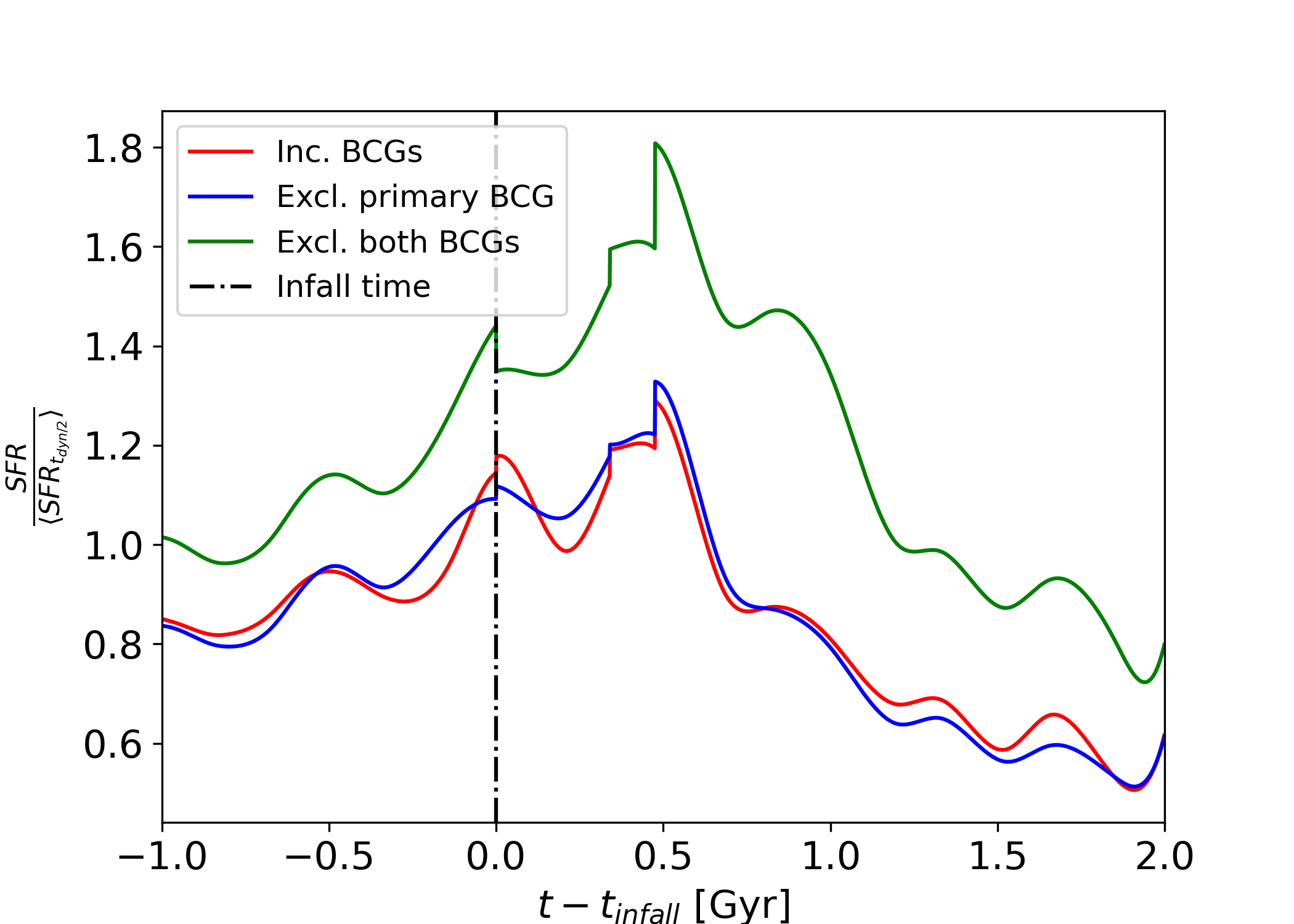}
		\caption{Average starbirth parameter, $b$, for selected cluster mergers as a function of time since infall, where infall is defined as the time corresponding to the peak of mass accretion. During the early stages of the merger -- up to approximately 500 Myr after infall -- the starbirth parameter exceeds unity, indicating a period of enhanced star formation. Following this phase, $b$ declines decreases even below one, consistent with a suppression of star formation activity.  }
		\label{Fig:Deltas}
\end{figure}

To assess the occurrence of starbursts during merging events, Figure~\ref{Fig:Deltas} presents the average birthrate parameter for merging clusters as a function of time since infall. On average, the results indicate an initial increase in star formation activity -- a starburst -- that lasts for approximately 500 Myr following infall, after which a rapid decline in the SFR occurs over the subsequent gigayear. This suggests that merger can initially trigger star formation through gas compression; however, over time, likely gas stripping and heating induced by the interaction act to quench star formation, as illustrated in Figure~\ref{Fig:evolution}. These interactions induce asymmetric gas distributions and the formation of gas tails in cluster galaxies, such as those shown in ~\ref{Fig:evolution}, which have been associated with enhanced star formation, driven by gas compression at the leading edge of galaxies~\citep{Safazadeh2019, Vulcani2020, Rohr2023}. 

These results agree with \citet{Fujita-Nagashima1999}, who studied the impact of ram pressure on the SFR of satellites in clusters. They found that the SFR increases by a factor of two when galaxies reach the first pericenter passage. However, after this point, there is a rapid quenching due to the effect of ram pressure near the central regions, resulting in the stripping of the cold gas and, thus, the depletion of molecular gas. This scenario for gas-rich merging clusters was also reported by~\cite{Stroe2014}, suggesting this short SFR increase could last just a few hundred million years. 
Our results complement those presented by~\citet{Contreras-Santos2022} who, using the Three Hundred Simulations, reported a slight increase in the young stellar population of BCGs following cluster mergers. They also found that merger-induced star formation was accompanied by an increase in luminosity and an excess of blue BCGs in the disturbed clusters compared to their relaxed counterpart.  Furthermore, our findings are consistent with the observational study by ~\citet{Stroe2014}, who, using optical spectroscopic data for the clusters CIZA J2242.8+5301 and 1RXS J0603.3+4213, showed that mergers in gas-rich clusters  can enhance SFRs. This enhancement is attributed to shock waves that compress the gas of the cluster galaxy members during these mergers. 

This initial enhancement in the SFR observed at the onset of the cluster interaction, followed by a decline as the system approaches a relaxed state, as well as, the physical processes responsible for this star formation enhancement and the subsequent quenching will be investigated and quantified in greater detail in a follow-up study.

\section{Summary and conclusions}

In this study, we have used the Illustris TNG100 simulations to investigate the relationship between the dynamic state of galaxy groups and clusters and the physical characteristics of their galaxy population. We selected a sample of groups and clusters with masses of $\text{M}_{200} \geq 10^{13}\; \text{M}_{\odot}$ and classified them between disturbed and relaxed based on the offset between the position of the BCG and the position of the centre-of-mass of the gas cells. Clusters were considered disturbed if this offset is larger than $0.4 \times \text{R}_{200}$. We identified 153 disturbed galaxy clusters within a redshift range between $0.1 < z \leq 1$  and compared them with an equivalent set of the 153 most relaxed galaxy clusters. We also divided our sample into high- and low-redshift structures to analyse the evolution of those properties. 

Galaxies in each sub-sample were classified as red and blue galaxies using a double Gaussian fit for the galaxies' colour distribution. The fits were performed for the high- and low-redshift groups and clusters.  We additionally separated quenched and star-forming galaxies following~\cite{Donnari2019}. Our analysis shows that the physical properties of satellite galaxies, such as colour, SFR, and gas availability, depend significantly on the relaxation state of their host clusters. Specifically, we found that clusters in a disturbed state have a higher fraction of blue galaxies and a lower fraction of quenched galaxies compared with their relaxed counterparts. We further find a significant dependence of the fraction of red galaxies and quenched galaxies with redshift of the cluster. We observed a higher proportion of blue galaxies and larger star formation activity at higher redshifts. The higher sSFR observed in galaxies bound to disturbed clusters aligns well with the study conducted by \cite{Cohen2015}. In their study, they analysed a sample of 379 galaxy clusters from the Sloan Digital Sky Survey, revealing an inverse correlation between the relaxation state of galaxy clusters and SFR. Those results are further supported by \citet{Pallero2022}. Using sample zoom-in simulations of galaxy clusters, they showed that, at low $z$,  a large fraction of the present-day cluster galaxy population arrives at the cluster pre-processed, regardless of the cluster mass.  However, at higher $z$  most galaxies reach their quenching state in situ, regardless of the cluster mass.  As a result, even massive clusters show a significant star-forming galaxy population that undergoes a rapid quenching phase after reaching the cluster $\text{R}_{200}$. Similarly, those findings are in agreement with ~\citet{Aldas2023}, who observed that in both high- and low-redshift, the galaxy population is more mixed in the disturbed clusters than in the relaxed ones, but the difference is more pronounced at high-redshifts.

We have also explored the differences in gas availability among galaxies bound to relaxed and disturbed clusters. The results show that galaxies in disturbed clusters consistently have higher quantities of gas in all phases - hot, diffuse, and condensed - compared to galaxies in relaxed clusters. The large amount of condensed gas can explain the higher star formation rate found in disturbed clusters and the lower fraction of quenched galaxies. The presence of more hot gas can be explained by the presence of shocks produced during the merging.   

Finally, in the specific case that the unrelaxed clusters correspond to merging systems, we found that during the initial phases of the merger, during up to the first ~ 500 Myrs after the infall, there is the presence of a starburst, likely due to gas compression resulting from the shocks generated during the merging process. As the galaxy cluster becomes more relaxed, star formation slows down, and galaxies become less active in forming new stars. These findings have significant implications for our understanding of galaxy evolution, highlighting the dynamic nature of star formation and the role of gas distribution in this process. To understand why this happens during mergers, we need to track and estimate the tidal forces and RPS acting on galaxies, which will be done in a follow-up work.

\begin{acknowledgements}
We thank the referee for their helpful comments, which improved the manuscript. FA was supported by the doctoral thesis scholarship of Agencia Nacional de Investigaci\'on y Desarrollo (ANID)-Chile, grant 21211648. FA acknowledges the Large Grant INAF 2023 - Witnessing the Birth of the Most Massive Structures of the Universe -  F.O. 1.05.23.01.03. FAG acknowledges funding from the Max Planck Society through a “PartnerGroup” grant. FAG acknowledges support from ANID FONDECYT Regular 1211370, the ANID Basal Project FB210003 and the HORIZON-MSCA-2021-SE-01 Research and Innovation Programme under the Marie Sklodowska-Curie grant agreement number 101086388. FAG acknowledges support from ANID Fondo 2021 GEMINI ASTRO21-0061. CVM acknowledges support from ESO Comité Mixto, and FONDEQUIP project EQM200216. The work of AZ and ERC is supported by NOIRLab, which is managed by the Association of Universities for Research in Astronomy (AURA) under a cooperative agreement with the U.S. National Science Foundation. ERC is supported by the International Gemini Observatory, a program of NSF NOIRLab, which is managed by the Association of Universities for Research in Astronomy (AURA) under a cooperative agreement with the U.S. National Science Foundation, on behalf of the Gemini partnership of Argentina, Brazil, Canada, Chile, the Republic of Korea, and the United States of America. Powered@NLHPC: This research was partially supported by the supercomputing infrastructure of the NLHPC (CCSS210001)
\end{acknowledgements}
\bibliographystyle{aa}
\bibliography{references}
\begin{appendix}
\onecolumn
\section{Gas Content excluding the second BCG}  
\begin{figure*}[h]
		\centering
		\includegraphics[width=\linewidth, trim={2.5cm, 2.0cm, 3cm, 3.5cm}, clip]{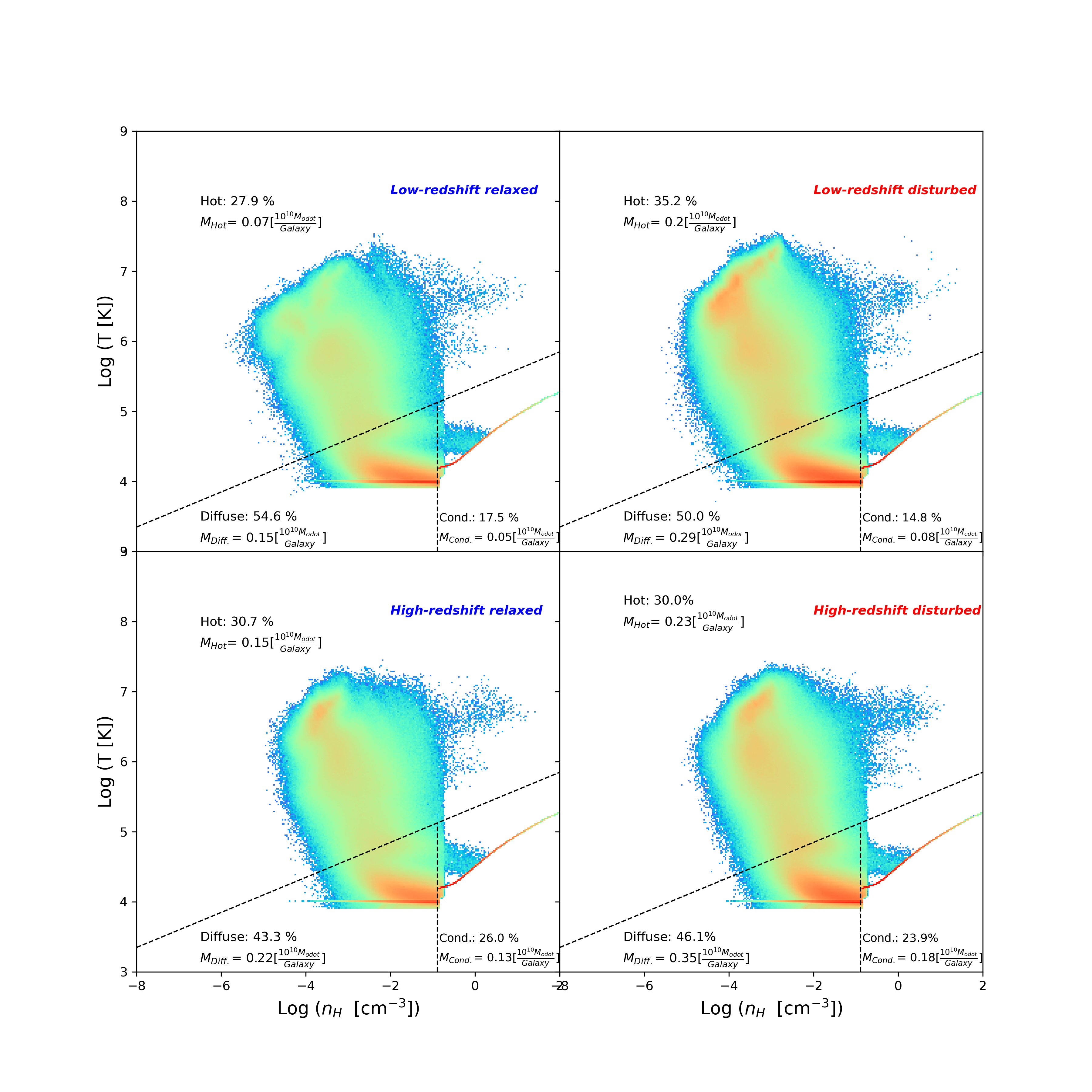}
		\caption{Gas content of the satellite galaxies in relaxed (left panels) and disturbed clusters (right panels), for low- (upper panels), and high-redshift (bottom panels). In those panels, in contrast to Figure \ref{Fig:GasTemp}, we exclude the second BCG in the case of disturbed halos. Each panel shows the gas content for each gas phase: hot, diffuse, and condensed. We can see that the hot gas content decreases, but we can see that on average galaxies in disturbed clusters contain more diffuse and condensed gas.   }
		\label{Fig:GasTempWOBCG}
\end{figure*}
\end{appendix}
\end{document}